\documentclass[preprintnumbers,superscriptaddress,nofootinbib,aps,prd,floatfix]{revtex4}

\usepackage{amsmath,slashed}
\usepackage{graphicx}

\newcommand{\be}{\begin{eqnarray*}}
\newcommand{\ee}{\end{eqnarray*}}

\newcommand{\bee}{\begin{eqnarray}}
\newcommand{\eee}{\end{eqnarray}}
\newcommand{\beeq}{\begin{equation}}
\newcommand{\eeeq}{\end{equation}}

\newcommand{\pt}{p_{\rm{T}}}

\newcommand{\ttbar}{\ensuremath{t\bar{t}}}

\def\spa#1.#2{\left\langle#1#2\right\rangle}
\def\spb#1.#2{\left[#1#2\right]}
\def\lor#1.#2{\left(#1#2\right)}
\def\sand#1.#2.#3{%
\left\langle\smash{#1}{\vphantom1}^{-}\right|{#2}%
\left|\smash{#3}{\vphantom1}^{-}\right\rangle}
\def\sandp#1.#2.#3{%
\left\langle\smash{#1}{\vphantom1}^{-}\right|{#2}%
\left|\smash{#3}{\vphantom1}^{+}\right\rangle}
\def\sandpp#1.#2.#3{%
\left\langle\smash{#1}{\vphantom1}^{+}\right|{#2}%
\left|\smash{#3}{\vphantom1}^{+}\right\rangle}
\def\sandpm#1.#2.#3{%
\left\langle\smash{#1}{\vphantom1}^{+}\right|{#2}%
\left|\smash{#3}{\vphantom1}^{-}\right\rangle}
\def\sandmp#1.#2.#3{%
\left\langle\smash{#1}{\vphantom1}^{-}\right|{#2}%
\left|\smash{#3}{\vphantom1}^{+}\right\rangle}
\def\spab#1.#2.#3{\langle#1|#2|#3]}
\def\spba#1.#2.#3{[#1|#2|#3\rangle}
\def\spaa#1.#2.#3{\langle#1|#2|#3\rangle}
\def\spbb#1.#2.#3{[#1|#2|#3]}
\def\spaxa#1.#2.#3.#4{\langle#1|#2|#3|#4\rangle}
\def\spbxb#1.#2.#3.#4{[#1|#2|#3|#4]}

\DeclareGraphicsExtensions{.pdf,.png,.jpg}

\usepackage{color}

\begin{document}

\title{Probing a light CP-odd scalar in di-top-associated production at the LHC}

\begin{abstract}
\noindent 
CP-odd scalars are an integral part of many extensions of the Standard Model. Recently, electroweak-scale pseudoscalars have received increased attention in explaining the diffuse gamma-ray excess from the Galactic Centre.
Elusive due to absence of direct couplings to gauge bosons, these particles receive only weak constraints from direct searches at LEP or searches performed during the first LHC runs. We investigate the LHC's sensitivity in observing a CP-odd scalar in di-top associated production in the mass range $20 \leq m_A \leq 100$ GeV using jet substructure based reconstruction techniques. We parametrise the scalar's interactions using a simplified model approach and relate the 
obtained upper limits on couplings within type-I and type-II 2HDMs as well as the NMSSM. We find that in di-top-associated production, experiments at the LHC can set tight limits on CP-odd scalars that fit the Galactic Centre excess.  However, direct sensitivity to light CP-odd scalars from the NMSSM proves to remain challenging.  
\end{abstract}

\author{Mirkoantonio Casolino}
\affiliation{Institut de F\'{\i}sica d'Altes Energies (IFAE), E-08193 Bellaterra, Barcelona, Spain}

\author{Trisha Farooque}
\affiliation{Institut de F\'{\i}sica d'Altes Energies (IFAE), E-08193 Bellaterra, Barcelona, Spain}

\author{Aurelio Juste}
\affiliation{Institut de F\'{\i}sica d'Altes Energies (IFAE), E-08193 Bellaterra, Barcelona, Spain}
\affiliation{Instituci\'o Catalana de Recerca i Estudis Avan\c{c}ats (ICREA), E-08010 Barcelona, Spain}

\author{Tao Liu}
\affiliation{Department of Physics, The Hong Kong University of Science and Technology, Clear Water Bay, Kowloon, Hong Kong S.A.R., P.R.C.}
\author{Michael Spannowsky}
\affiliation{Institute for Particle Physics Phenomenology, Department of Physics,\\Durham University, DH1 3LE, United Kingdom}

\preprint{IPPP/15/37}
\preprint{DCPT/15/74}

\maketitle

\section{Introduction}
\label{sec:intro}
The recent discovery of the Higgs boson \cite{Chatrchyan:2012ufa,Aad:2012tfa} marked a new era for fundamental physics. For the first time an electroweak-scale scalar resonance has been discovered, supposedly a remnant of the mechanism underlying electroweak symmetry breaking \cite{orig}.

While elementary scalar particles have been observed in nature for the first time, they are often an integral part of Standard Model (SM) extensions, e.g. Supersymmetry or general N-Higgs Doublet Models. When these extensions contain complex scalar fields, as a result, CP-odd scalars are introduced to the spectrum of the theory. Hence, since their existence would be evidence for physics beyond the SM, searches for CP-odd scalars are at the core of the 
current LHC program. 

Recently, CP-odd scalars as mediators between Dark Matter (DM) and SM particles have received attention in explaining the diffuse gamma-ray excess from the Galactic Centre~\cite{Goodenough:2009gk,Hooper:2010mq,Abazajian:2012pn,Daylan:2014rsa} in the contexts of the so-called Coy Dark Matter models~\cite{Boehm:2014hva,Hektor:2014kga,Arina:2014yna}, and the Next-to-Minimal-Supersymmetric-Standard-Model (NMSSM)~~\cite{Cheung:2014lqa,Huang:2014cla}. Hence, they are included as mediators in simplified models by the ATLAS and CMS collaborations to recast searches for  jets and missing transverse energy (`monojet') during the upcoming LHC runs \cite{Malik:2014ggr,Abdallah:2014hon}.

In contrast to the widely accepted paradigm that new physics particles have to be heavy, i.e. masses of $\mathcal{O}$(1) TeV or beyond, the mass of CP-odd scalars is almost unconstrained by direct searches. As interactions between gauge bosons and CP-odd scalars are only induced via higher-dimensional operators, e.g. $\frac{1}{2} A \epsilon_{\mu \nu \sigma \rho} V_{\sigma \rho} V^{\mu \nu}$, limits from LEP are fairly weak. The main collider sensitivities may mainly arise from bottom quark or top quark-associated productions (for recent explorations, see~\cite{Craig:2015jba,Hajer:2015gka}). Further, due to the predicted velocity suppression in direct detection experiments for CP-odd scalar mediators, even light CP-odd scalars are still in agreement with experimental observations. Some constraints from flavour physics exist but limits are again weak if $m_A \gtrsim 5$ GeV \cite{Dolan:2014ska}, assuming the CP-odd scalar interacts with fermions in agreement with the hypothesis of minimal flavour violation \cite{D'Ambrosio:2002ex}. 

Therefore, indirect detection experiments and direct searches at the LHC appear to be the most sensitive ways to search for the existence of electroweak-scale CP-odd scalar particles. In this paper, instead of previously explored paths of searching for CP-odd scalars in gluon-fusion production \cite{Klamke:2007cu,Dolan:2014upa}, we focus on the direct production of such particles in association with a top quark pair and subsequent decay into a bottom quark pair, $p p \to t \bar{t} A \to t \bar{t} b \bar{b}$. Thus, we derive limits on the mass and coupling strength of the CP-odd scalar in a process with unsuppressed fermion couplings only.\footnote{We note that, if the pseudoscalar couples for example in a universal way to fermions as part of a UV-complete model, thereby not respecting Yukawa-like coupling hierarchies, other production and decay channels might be more sensitive. However, the analysis we provide is still valid as a subset of possible search channels.}

This paper is organised as follows. In Sec.~\ref{sec:model} we briefly outline the way we incorporate the CP-odd scalar into the theory, using a simplified model approach. The event generation and details of the final state reconstruction are described in detail in Secs.~\ref{sec:generation} and \ref{sec:analysis}. In Sec.~\ref{sec:limit} we derive limits on the mass of the CP-odd scalar and its couplings to top quarks. Such limits can be applied to models where the CP-odd scalar arises as part of a Higgs multiplet. We recast these limits in the context of the 2HDM and the NMSSM in Sec.\ref{sec:interpretation}. Finally, in Sec.\ref{sec:conclusion} we offer conclusions.

\section{Simplified model}
\label{sec:model}
While CP-odd scalars are present in many extensions of the SM, for simplicity and generality of our results, we use a simplified model approach \cite{Alves:2011wf} to parametrise the contribution of this particle in the process $p p \to t \bar{t} A \to t \bar{t} b \bar{b}$. More precisely, we add couplings of the CP-odd scalar with the bottom and top quarks to the full SM Lagrangian 
\begin{eqnarray}
\mathcal{L} = \mathcal{L}_{\mathrm{SM}} + \mathcal{L}_{\mathrm{CP-odd}},
\label{eq:lagrangian}
\end{eqnarray}
where 
\begin{eqnarray}
\mathcal{L}_{\mathrm{CP-odd}} = i \frac{g_{t} y_t}{\sqrt{2}}~\bar{t} \gamma_5 t A + i \frac{g_{b} y_b}{\sqrt{2}}~\bar{b} \gamma_5 b A,
\end{eqnarray}
and $g_i$ ($i=t,b$) parametrises the deviation from the SM Yukawa coupling  $y_i = m_i/v$.

Recently, a similar approach was proposed to recast monojet searches at the LHC in terms of scalar mediators between the SM and a secluded sector \cite{Buckley:2014fba,Harris:2014hga,Haisch:2015ioa,Khoze:2015sra}. In a similar way, we will focus on the minimal set of free parameters relevant to the process considered. Throughout this paper we will assume $A$ to be a narrow resonance with $2 m_b \leq m_A < 2m_t$. Hence, in our approach the CP-odd scalar decays exclusively into bottom quarks with ${\cal B}(A\to b\bar{b})=1$ and its width $\Gamma_A$ is completely determined by the value of $g_b$. 
For a narrow resonance, the kinematic distributions are expected to remain largely independent of the value of $\Gamma_A$.

\section{Event generation and simulation details}
\label{sec:generation}

\subsection{Signal and background modelling}

Signal and background samples corresponding to $pp$ collisions at $\sqrt{s}=14$~TeV are generated using 
the {\sc Madgraph5} 2.1.1~\cite{Alwall:2011uj} leading-order (LO) generator and the  CTEQ6L1~\cite{Nadolsky:2008zw}  
set of parton distribution functions (PDF), interfaced to {\sc Pythia} v6.427~\cite{Sjostrand:2006za}  for parton 
showering and fragmentation and using the Perugia2011C~\cite{Skands:2010ak}  underlying event tune. In all cases, a top quark mass 
of 172 GeV is assumed and top quarks are decayed inclusively by {\sc Pythia}. 

\begin{figure}[htbp]
\begin{center}
\begin{tabular}{cc}
\includegraphics[width=0.4\textwidth]{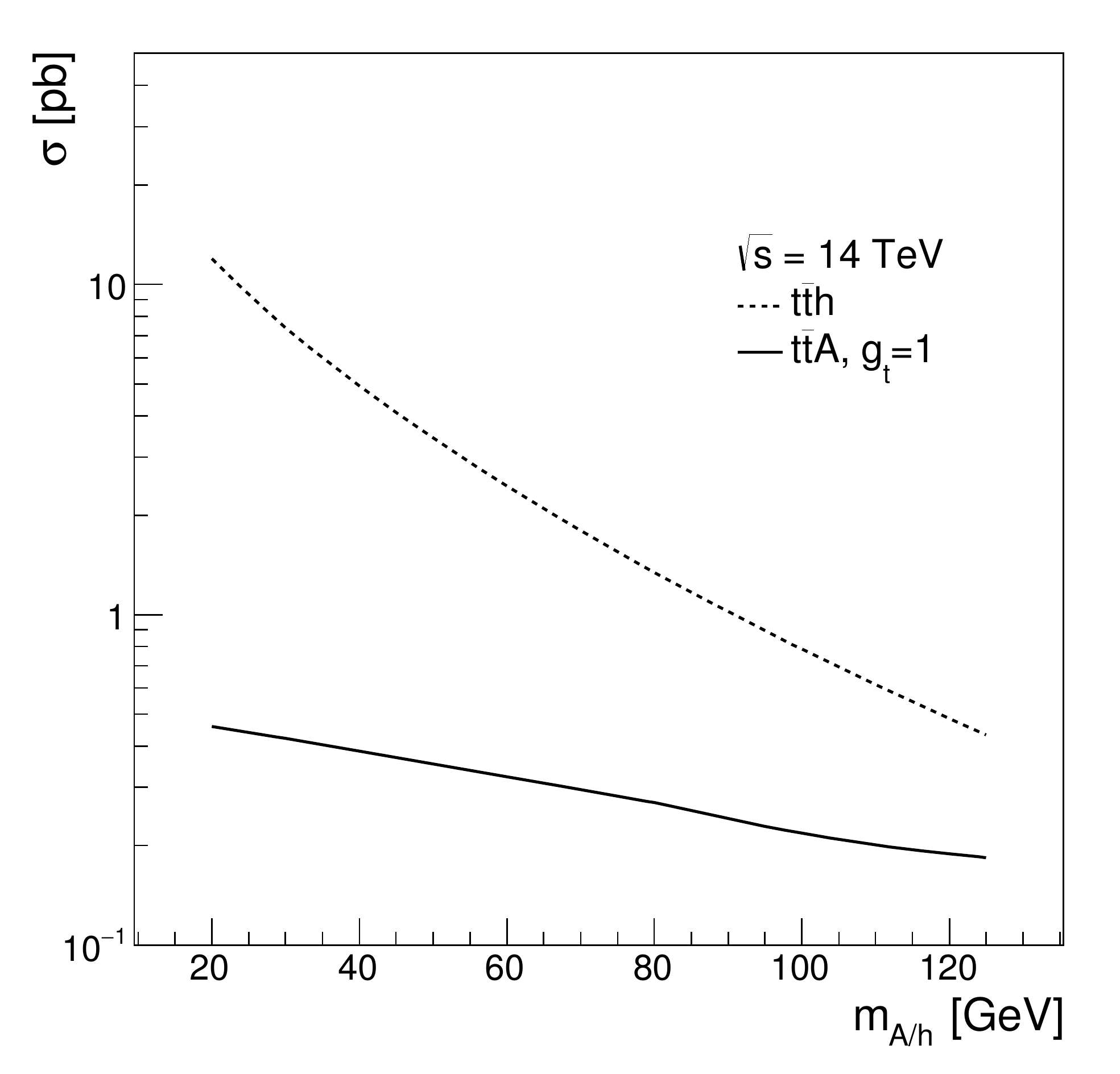} &
\includegraphics[width=0.4\textwidth]{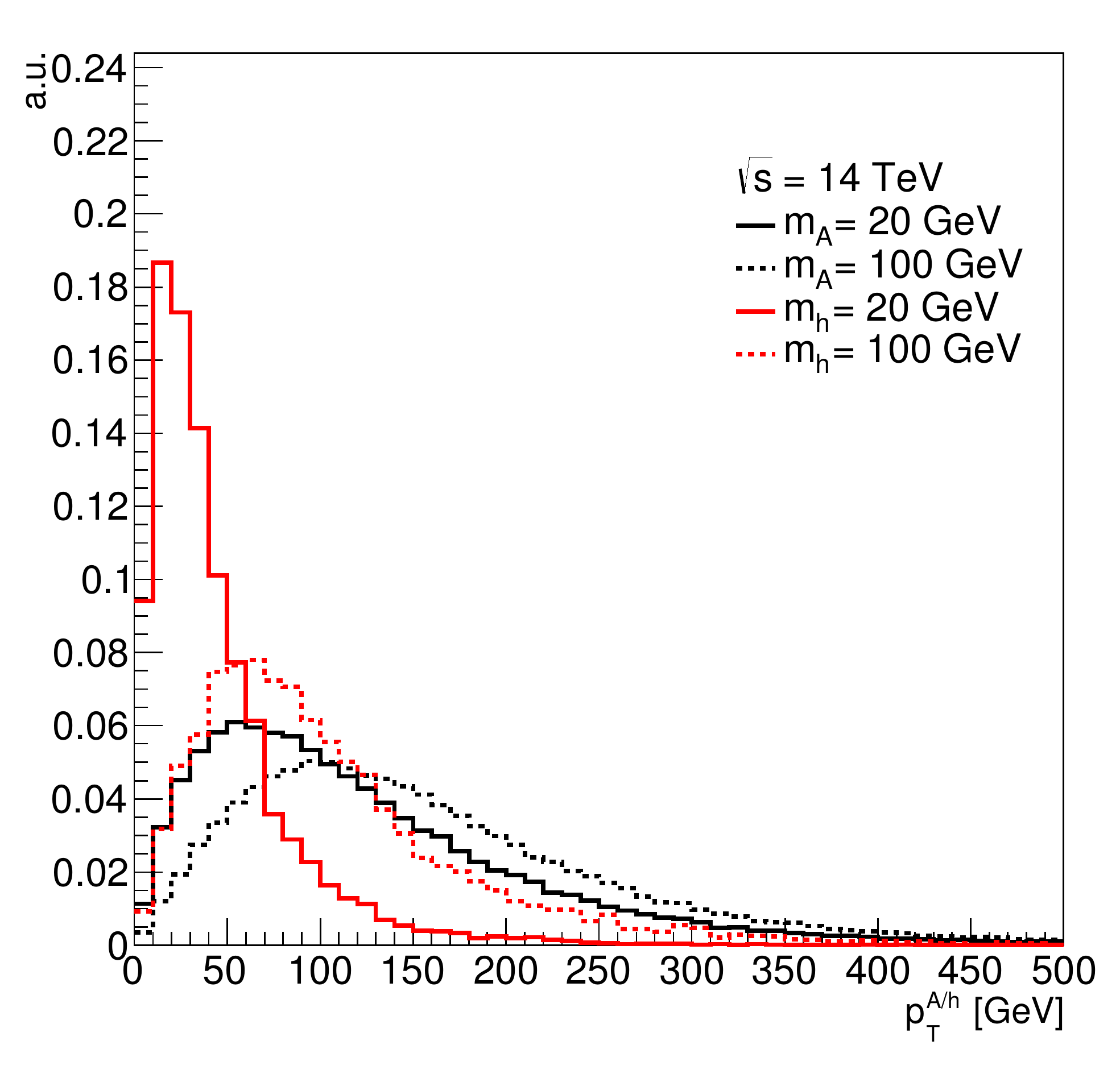} \\
(a) & (b) \\
\end{tabular}
\caption{\small {(a) Comparison of the leading-order cross section for $\ttbar h$ (solid line) and 
$\ttbar A$ (dashed line) in $pp$ collisions at $\sqrt{s}=14$ TeV as a function of Higgs boson mass.
In both cases a value of $g_t = 1$ is assumed. (b) Comparison of the Higgs boson $\pt$ between 
$\ttbar h$ (red) and $\ttbar A$ (black) for two different values of the Higgs boson mass, 20 GeV (solid) and 100 GeV (dashed).}}
\label{fig:xsect_pTA_ttA} 
\end{center}
\end{figure}

Samples of $\ttbar A$ signal events are generated for different values of the $A$ boson mass, $m_A = 20, 30,40, 60, 80$ and 100 GeV,
and assuming $g_t=1$ and ${\cal B}(A \to b\bar{b})=1$. A model corresponding to the Lagrangian shown in Eq.~\ref{eq:lagrangian} 
is implemented using Feynrules 2.1~\cite{Alloul:2013bka} and imported as UFO  model~\cite{Degrande:2011ua} in {\sc Madgraph5}.  
The LO signal cross section predicted by  {\sc Madgraph5} (see Table~\ref{tab:sigma_ttA}) is scaled by a k-factor of 1.3. This k-factor is obtained 
as the ratio of the NLO to LO cross sections for $\ttbar h$ production, where $h$ is a CP-even Higgs boson. It has been checked that this k-factor
is rather constant as a function of $m_h$, varied between 20 and 125 GeV.
Figure~\ref{fig:xsect_pTA_ttA}(a) compares the production cross section between $\ttbar h$ and $\ttbar A$ as a function of the Higgs boson
mass, in both cases assuming $g_t$=1. The ratio between both cross sections varies significantly versus mass, with the $\ttbar h$ cross section
being about a factor of 20 larger than the $\ttbar A$ cross section at a mass of 20 GeV, and only about a factor of two larger at a mass of 120 GeV \cite{Frederix:2011zi}.
This difference results from the presence of the extra $\gamma_5$ factor in the interaction between a CP-odd Higgs boson and the top quark,
compared to the case of a CP-even Higgs boson.
Another consequence of the different interaction is that a CP-odd Higgs boson has a  substantially harder $\pt$ spectrum compared to the  CP-even case,
particularly at low mass, as illustrated in Fig.~\ref{fig:xsect_pTA_ttA}(b). This is a key feature exploited in this analysis, as discussed in
Sec.~\ref{sec:analysis}.

\begin{table}[h] 
\begin{center} 
\begin{tabular}{ccccccc} 
\hline\hline
$\quad$ $m_A$ (GeV) $\quad$ & $\quad$ 20 $\quad$ & $\quad$ 30 $\quad$ & $\quad$ 40 $\quad$ & $\quad$ 60 $\quad$ & $\quad$ 80 $\quad$ & $\quad$ 100 $\quad$ \\
\hline
$\sigma^{\rm LO}(\ttbar A)$ (pb) & 0.46 & 0.42 & 0.39 & 0.32 & 0.27 & 0.23 \\
\hline\hline
\end{tabular} 
\caption{\small {Leading-order cross section for $\ttbar A$ production in $pp$ collisions at $\sqrt{s}=14$ TeV as a function of 
the $A$ boson mass $m_A$. As discussed in the text, this LO cross section is obtained assuming $g_t=1$ and will be 
multiplied by a k-factor of 1.3 to approximate the NLO cross section.}}
\label{tab:sigma_ttA} 
\end{center} 
\end{table} 

A large sample of $\ttbar$+jets background events is generated with up to two additional partons  in the 5F scheme 
(i.e. including $b$- and $c$-quarks). To avoid double-counting of partonic configurations generated by both the 
matrix-element  calculation and the parton shower,  a parton--jet matching scheme (``MLM matching'')~\cite{Mangano:2001xp} 
is employed. The sample is normalised to a cross section of 990 pb obtained using {\sc Top++} v2.0~\cite{Czakon:2011xx} 
at next-to-next-to-leading order (NNLO) in QCD, including resummation of next-to-next-to-leading logarithmic (NNLL) soft gluon 
terms~\cite{Cacciari:2011hy,Baernreuther:2012ws,Czakon:2012zr,Czakon:2012pz,Czakon:2013goa}, 
and using the MSTW 2008 NNLO~\cite{Martin:2009iq,Martin:2009bu} PDF set.
The $\ttbar$+jets sample is generated inclusively, but events are categorised depending
on the flavour content of additional particle jets in the event (i.e. jets not originating from
the decay of the $\ttbar$ system). Particle jets are reconstructed with the anti-$k_t$~\cite{Cacciari:2008gp,Cacciari:2005hq,Cacciari:2011ma} 
algorithm with a radius parameter $R=0.4$ and are required to have $\pt>15$~GeV and
$|\eta|<2.5$. Events where at least one such particle jet is matched within $\Delta R<0.4$ to a $b$-hadron
with $\pt>5$~GeV not originating from a top quark decay are generically labelled as $\ttbar$+$\geq$$1b$ events.
Similarly, events where at least one such particle jet is matched within $\Delta R<0.4$ to a $c$-hadron
with $\pt>5$~GeV not originating from a $W$ boson decay, and that are not labelled already as $\ttbar$+$\geq$$1b$, 
are labelled as $\ttbar$+$\geq$$1c$  events. Events labelled as either $\ttbar$+$\geq$$1b$  or
$\ttbar$+$\geq$$1c$ are generically referred to below as $\ttbar$+HF events, where HF stands for ``heavy flavour''.
We do not apply dedicated corrections to the normalisation of $\ttbar$+HF events, since Run 1 searches at the LHC~\cite{Aad:2015gra} 
showed that the LO prediction from {\sc Madgraph5} using the same settings as us is consistent with data within $\sim 20\%$,
and a larger systematic uncertainty will be assumed in this study.
As in Ref.~\cite{Aad:2015gra}, a finer categorisation of $\ttbar$+HF events is considered for the purpose of assigning systematic uncertainties
associated with the modelling of heavy-flavour production in different topologies. In this way, a distinction is
made between events with only one extra heavy-flavour jet satisfying the above cuts (referred to as $\ttbar$+$b$ or $\ttbar$+$c$),
events with two extra heavy-flavour jets (referred to as $\ttbar$+$b\bar{b}$ or $\ttbar$+$c\bar{c}$), and events
with one extra heavy-flavour jet containing two $b$- or $c$-hadrons  (referred to as $\ttbar$+$B$ or $\ttbar$+$C$).
The remaining events are labelled as $\ttbar$+light-jet events, including those with no additional jets. 

Additional background samples corresponding to $\ttbar W$, $\ttbar Z$ and $\ttbar h_{\rm SM}$ production, where $h_{\rm SM}$ is the SM Higgs boson,
are also produced. The $\ttbar W$ sample is generated requiring at least one $W$ boson in the event to decay leptonically,
and is normalised to the corresponding LO cross section, 0.404 pb, times a k-factor of 1.4~\cite{Garzelli:2012bn}.
The $\ttbar Z$ sample is generated requiring $Z \to q\bar{q}$ decays and is normalised
to the corresponding LO cross section, 0.353 pb, times a k-factor of 1.3~\cite{Garzelli:2012bn}.
Finally, the $\ttbar h_{\rm SM}$ sample is generated assuming $m_h=125$~GeV and requiring $h \to b\bar{b}$ decays.
It is normalised to the NLO cross section~\cite{Dawson:2003zu,Beenakker:2002nc,Beenakker:2001rj}, 0.611 pb, 
times the $h_{\rm SM} \to b\bar{b}$ branching ratio of 57.7\%~\cite{Djouadi:1997yw,Bredenstein:2006rh,Actis:2008ts,Denner:2011mq},
collected in Ref.~\cite{Dittmaier:2011ti}.
In these samples $Z \to q\bar{q}$ and $h_{\rm SM} \to b\bar{b}$ decays are performed by {\sc Madgraph5} and top quarks 
and $W$ bosons are decayed by {\sc Pythia}. 

\subsection{Event reconstruction}
\label{sec:event_reco}

The generated  samples at the particle level are processed through a simplified simulation of the detector
response and object reconstruction.

Isolated leptons (electrons or muons) are required to originate from a $W$ boson or $\tau$-lepton decay and
to have $\pt>25$ GeV and $|\eta|<2.5$. Furthermore, they are required to not overlap with jets, as discussed
below. A typical per-lepton identification efficiency of 80\% is assumed.

Stable particles from {\sc Pythia}, except for muons and neutrinos, are processed through a simplified
simulation of a calorimeter. The four momenta of particles falling within the same window in $\eta$--$\phi$ space of 
size $\Delta\eta \times \Delta\phi = 0.1\times 0.1$ are added together to simulate the finite granularity of
calorimeter cells. For each cell, the total three momentum is rescaled such as to make the cell massless.
Cells with energy larger than 0.1 GeV and $|\eta|<5.0$ become the inputs to the jet algorithm.
Several types of jets are considered in this analysis.

The anti-$k_t$ algorithm is used to reconstruct jets with two different radius parameters, $R=0.2$ and $R=0.4$, referred
to as AKT2 and AKT4 jets respectively. The minimum jet $\pt$ threshold for reconstruction is 5 GeV. 
During jet reconstruction, no distinction is made between identified electrons and jet energy deposits, and
so every electron is also reconstructed a jet. In order to remove this double counting, if any of the jets in the AKT2 and AKT4 collections
lie within $\Delta R=0.2$ of a selected electron, the closest jet from each jet collection is discarded.
Since this analysis has a large number of $b$-quark initiated jets, for which a significant fraction of energy
is carried away by muons in semi-muonic $b$-hadron decays, the four momenta of all reconstructed muons 
with $\pt>4$ GeV that are ghost-associated~\cite{Cacciari:2007fd,Cacciari:2008gn} to a jet are added to the calorimeter jet four momentum.
After this correction, a minimum $\pt$ requirement of 15 GeV and 25 GeV is made
for AKT2 and AKT4 jets respectively. All jets are required to satisfy $|\eta|<2.5$. 
Finally, any electron or muon within $\Delta R=0.4$ of a selected AKT4 jet is discarded.
In this analysis AKT4 jets are used to define the minimum jet multiplicity required in the event selection, while 
AKT2 jets are used to define the $b$-tag multiplicity of the event. The latter is particularly important since 
at low $m_A$ values the $b$-quarks from the $A \to b\bar{b}$ decay emerge with small angular separation.
The flavour of an AKT2 jet is determined by matching it within $\Delta R=0.15$ with a $b$-hadron or 
a $c$-hadron (not originating from a $b$-hadron decay), resulting in the jet being labelled as $b$-jet or $c$-jet respectively. 
The rest of the jets are taken to originate from the fragmentation of a light quark or gluon and are labelled 
as ``light jets". Heavy-flavour tagging is modelled in a probabilistic fashion by assigning a per-jet efficiency of 
70\% to $b$-jets, 20\% to $c$-jets, and 0.7\% to light jets.

In addition, jets are reconstructed with the Cambridge-Aachen (C/A) algorithm~\cite{Dokshitzer:1997in,Wobisch:1998wt} 
in order to reconstruct the $A \to b\bar{b}$ decay, taking advantage of the boost with which $A$ bosons are produced in the
$\ttbar A$ process. Two radius parameters are considered for C/A jets, $R^{\rm C/A}=0.6$ and 0.8, referred to as CA6 and CA8 jets 
respectively. The choice of radius for C/A jets is optimised in order to optimally
reconstruct the $\ttbar A$ signal depending on the value of $m_A$. In order to minimise the impact of
soft radiation and pileup (not modelled in this analysis), the mass-drop (a.k.a. BDRS) filtering algorithm~\cite{Butterworth:2008iy, Plehn:2009rk} 
with the following parameters, $\mu_{\rm frac}=0.67$ and $y_{\rm cut}=0.09$~\cite{Aad:2013gja}, is applied to the
reconstructed C/A jets. A semi-muonic energy correction to the C/A jet four momentum is also applied, as in the case of AKT2 and AKT4 jets.

\section{Experimental analysis}
\label{sec:analysis}

\subsection{Analysis strategy and event selection}
\label{subsec:analysis}
This search is focused on the $\ttbar A \to W^+b W^- \bar{b}b\bar{b}$ process, with one of the $W$ bosons
decaying leptonically and the other $W$ boson decaying hadronically. Only electrons or muons originating 
from $W$ boson or $\tau$-lepton decays are considered. The resulting final state signature is thus characterised
by one electron or muon, and high jet and $b$-jet multiplicity that can be exploited to suppress the background,
dominated by $\ttbar$+jets production. Therefore, the following preselection requirements are made:
one electron or muon, $\geq$5 AKT4 jets and $\geq$3 AKT2 $b$-tagged jets, in the following simply referred
to as $\geq$5 jets and $\geq$3 $b$-tags. In order to optimise the sensitivity of the search, the selected events 
are categorised into two separate channels depending on the number of $b$-tags (3 and $\geq$4).
The channel with $\geq$5 jets and $\geq 4$ $b$-tags has the largest signal-to-background ratio and 
therefore drives the sensitivity of the search. It is dominated by $\ttbar$+HF background. The channel with 3 $b$-tags 
has significantly lower signal-to-background ratio and the background is enriched in $\ttbar$+light-jets.
The simultaneous analysis of both channels is particularly useful to calibrate {\sl in-situ} the $\ttbar$+jets background prediction 
(including its heavy-flavour content) and constrain the related systematic uncertainties, as it will be discussed in
Sec.~\ref{sec:stat_analysis}. This is a common strategy used in many experimental searches in the 
ATLAS and CMS collaborations~\cite{Aad:2015gra,Khachatryan:2015ila,Aad:2015kqa}, 
which we mimic here in order to obtain more realistic projected sensitivities.

\begin{figure}[htbp]
\begin{center}
\begin{tabular}{cc}
\includegraphics[width=0.4\textwidth]{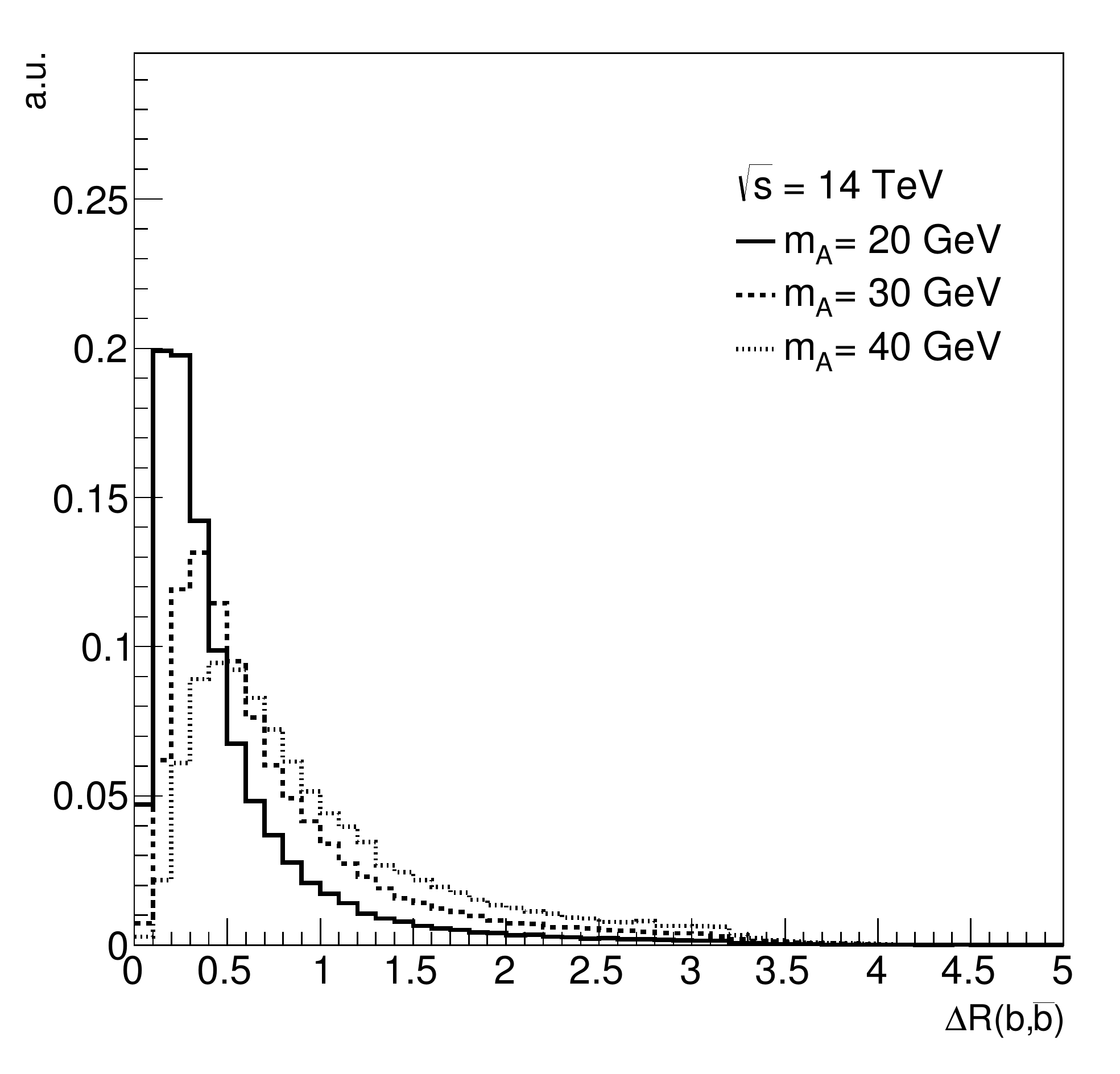} &
\includegraphics[width=0.4\textwidth]{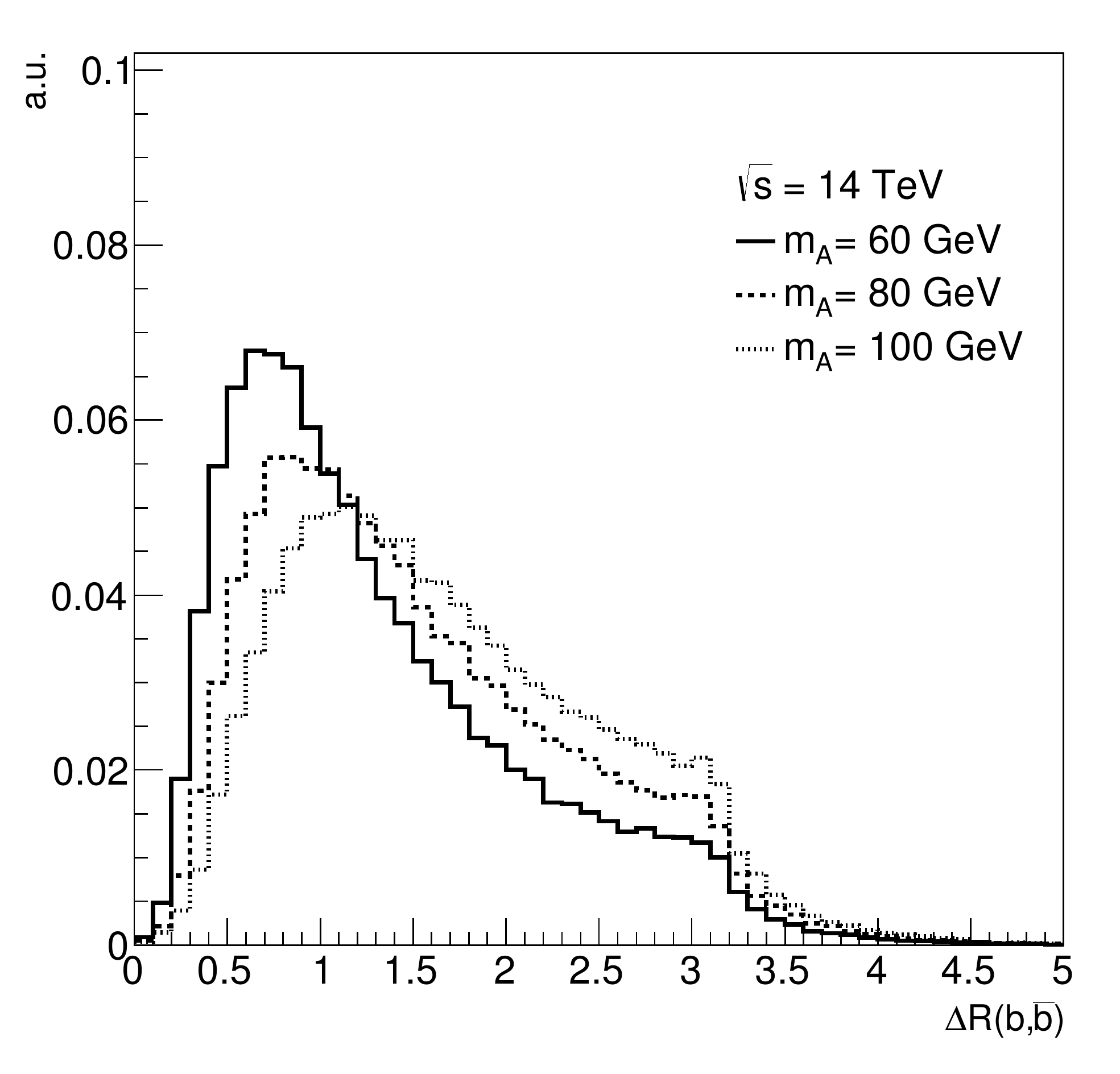} \\
(a) & (b) \\
\end{tabular}
\caption{\small {Distribution of $\Delta R$ between the two $b$-quarks from the $A \to b\bar{b}$ decay prior to
any selection requirements, for different values of $m_A$: (a) $m_A=20, 30$ and  40 GeV, and (b) $m_A=60, 80$ and 100 GeV.}}
\label{fig:DRbb_ttA} 
\end{center}
\end{figure}

An extra handle is provided by the significant boost of the $A$ boson in a fraction of signal 
events, which results in the two $b$-jets from the $A \to b\bar{b}$ decay emerging with small angular separation between them.
This is particularly relevant for low $m_A$ values, as shown in Fig.~\ref{fig:DRbb_ttA}.
As a result, the $A$ boson decay products can be reconstructed into a single fat jet, whose mass distribution would show a 
resonant structure peaked at the correct $m_A$ value. This feature is also very powerful to discriminate against the background.
Therefore, a further requirement is made to have at least one C/A BDRS-filtered jet with radius parameter $R^{\rm CA}$ and minimum $\pt$
depending on the $m_A$ hypothesis being tested. In order to correctly reconstruct a significant fraction of the signal while rejecting as 
much background as possible,  CA6 jets are used for $m_A \leq 40$ GeV, while CA8 jets are used for higher $m_A$ values (up to 100 GeV). 
The minimum $\pt$ requirements on the C/A jets are 60, 100, 120, 150, 200 and 250 GeV for
$m_A=20, 30, 40, 60, 80$ and 100 GeV, respectively. 
As shown in Fig.~\ref{fig:DRbb_ttA}, for high values of $m_A$ only a small fraction of signal events would have the $A$ decay products contained
within the CA8 jet. The small signal acceptance comes with the benefit of improved background rejection and
the ability to reconstruct the $A$ boson mass, desirable in such simple analysis. However, it is expected that a dedicated multivariate 
analysis focused on the sample rejected by this analysis, similar in spirit to the ATLAS and CMS searches for the SM Higgs boson in 
$t\bar{t} h$, $h \to b\bar{b}$~\cite{Aad:2015gra,Khachatryan:2015ila}, could also achieve significant signal sensitivity at high $m_A$. 
Evaluating this possibility is beyond the scope of this study.
The number of $b$-tags inside the C/A jet is determined by
matching the $b$-tagged AKT2 jets within a cone of radius $\Delta R =0.75 R^{\rm C/A}$. 
Finally, a requirement is made is that the C/A jets have $\geq$ 2 $b$-tags inside. In case of more than one selected C/A jet,
the leading $\pt$ one is chosen. 

Table~\ref{tab:yields} presents the expected yields for signal and the SM backgrounds 
per fb$^{-1}$ of integrated luminosity as a function of the selection cuts applied in each
of the analysis channels under consideration: ($\geq$5j, 3b) and ($\geq$5j, $\geq$4b).
In the case of the ($\geq$5j, 3b) channel, the dominant background after final selection
is $\ttbar$+light-jets, where typically the two $b$-quarks from the  top quark decays, as well as
the $c$-quark from the $W \to c\bar{s}$ decay, are $b$-tagged.
In contrast, in the ($\geq$5 j, $\geq$4 b) channel half of the background is $\ttbar$+$\geq$$1b$,
with $\ttbar$+$b\bar{b}$ being its leading contribution. The rest of the background is approximately 
equally split between $\ttbar$+$\geq$$1c$ and $\ttbar$+light-jets.
In this table the expected contribution from $\ttbar A$ signal is obtained under the assumptions of $g_t=2$ and ${\cal B}(A\to b\bar{b})=1$.
Both analysis channels have approximately the same amount of signal, while the background
is about a factor of four higher in the ($\geq$5j, 3b) channel than in the  ($\geq$5j, $\geq$4b) channel.
Together with the different composition of the background, the very different signal-to-background
ratio between both channels is the primary motivation for analysing them separately.

\begin{figure}[htbp]
\begin{center}
\begin{tabular}{ccc}
$m_A = 20$~GeV & $m_A = 30$~GeV &  $m_A = 40$~GeV \\
\includegraphics[width=0.3\textwidth]{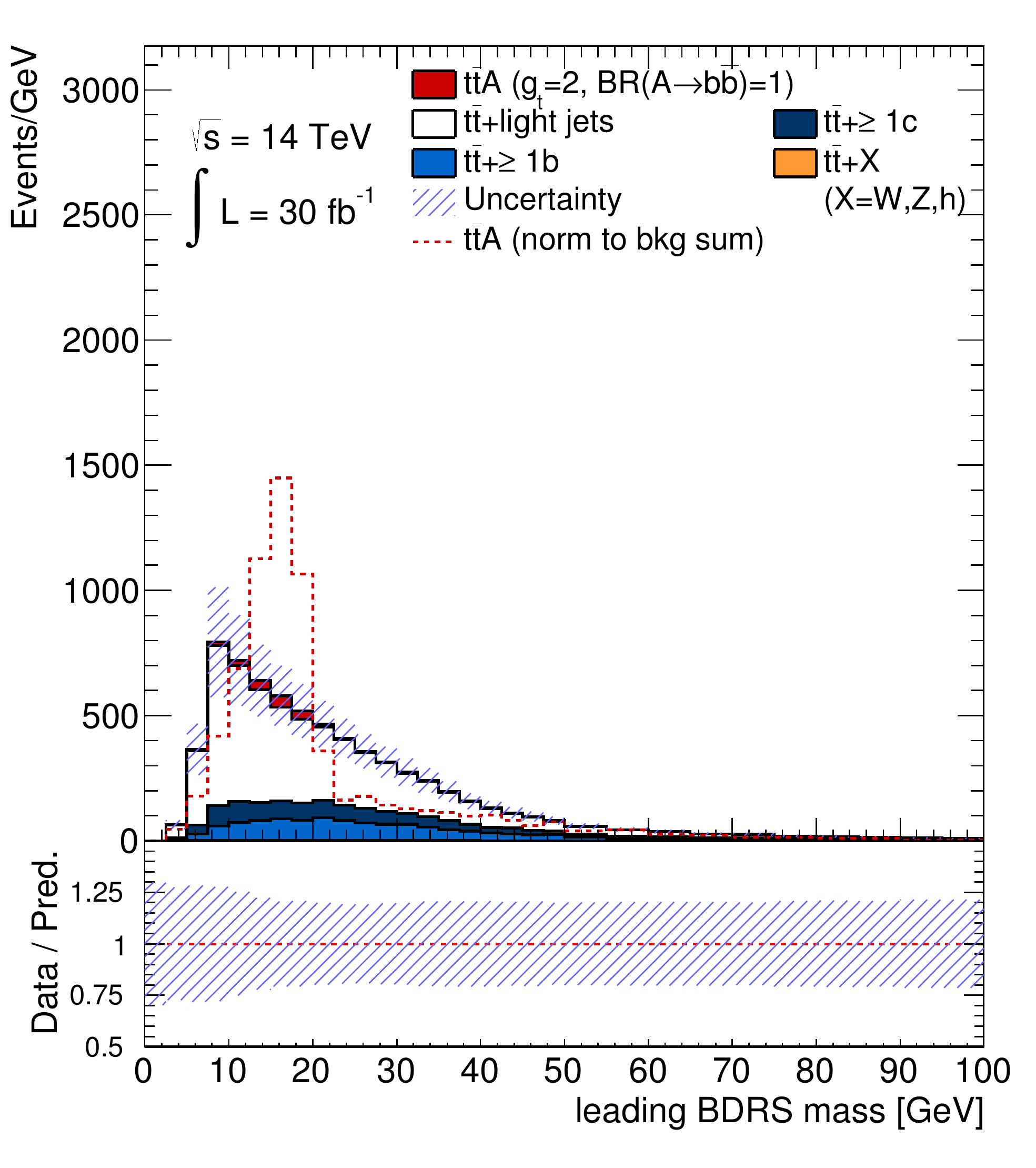} &
\includegraphics[width=0.3\textwidth]{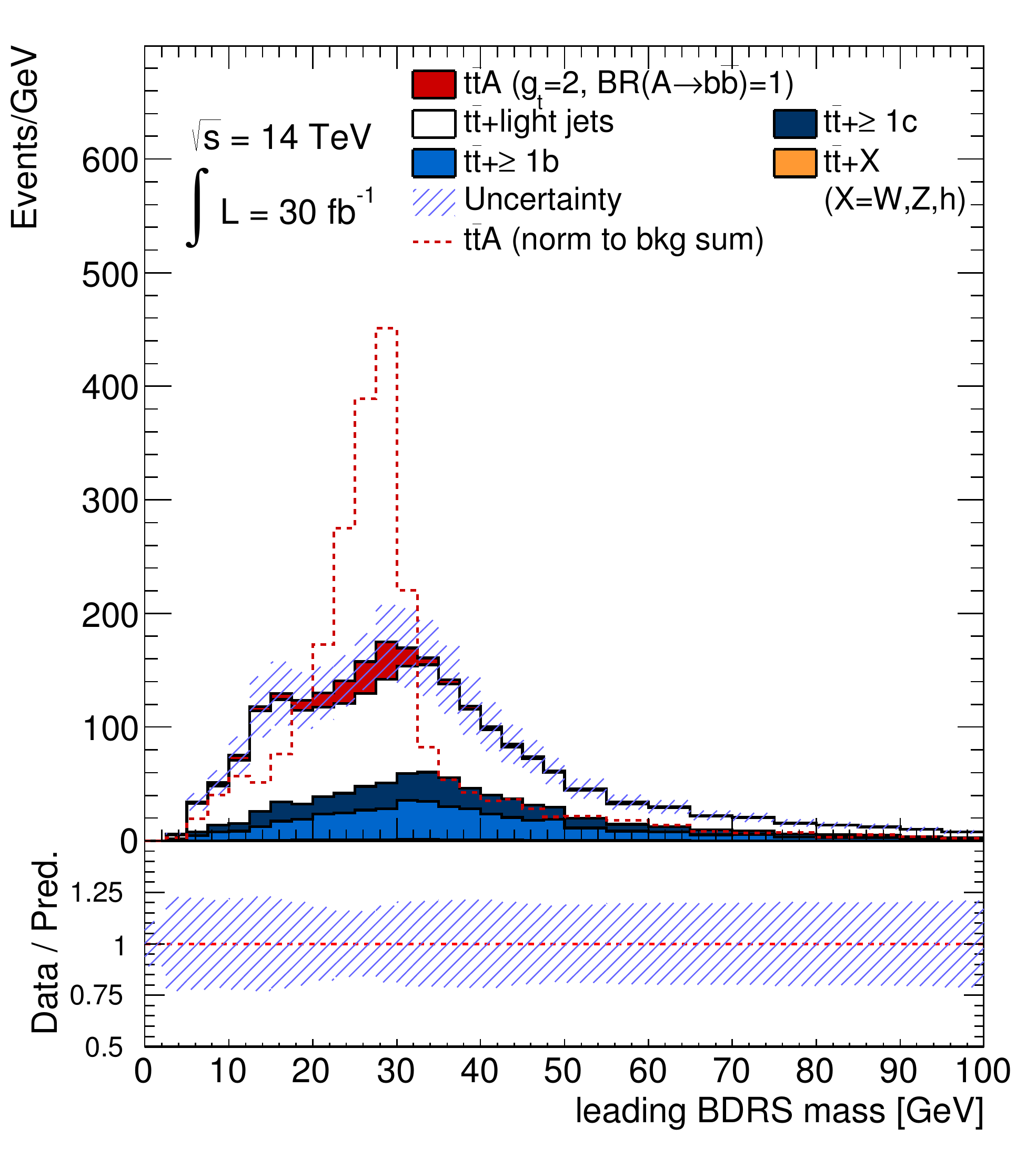} &
\includegraphics[width=0.3\textwidth]{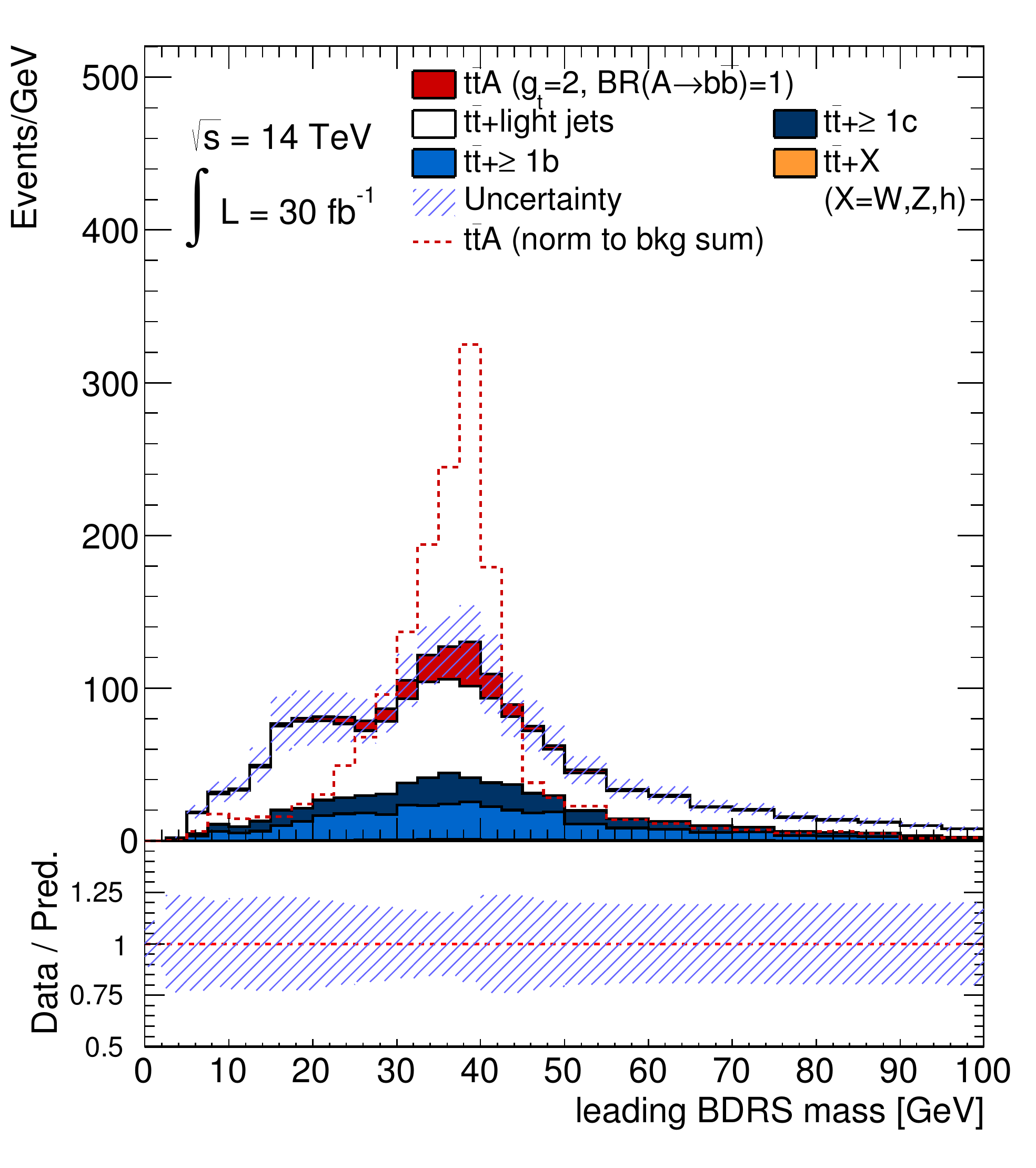} \\
\includegraphics[width=0.3\textwidth]{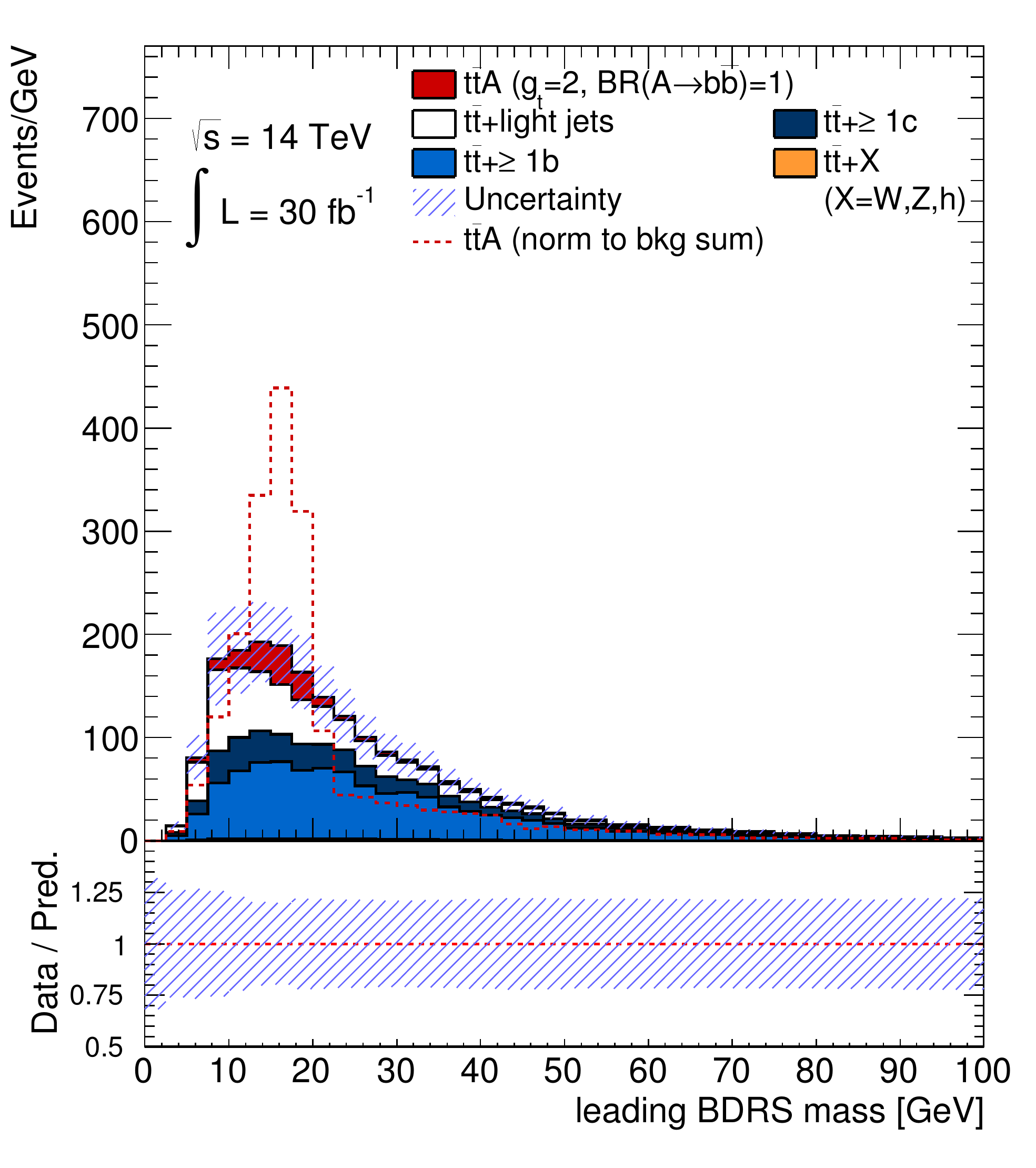} &
\includegraphics[width=0.3\textwidth]{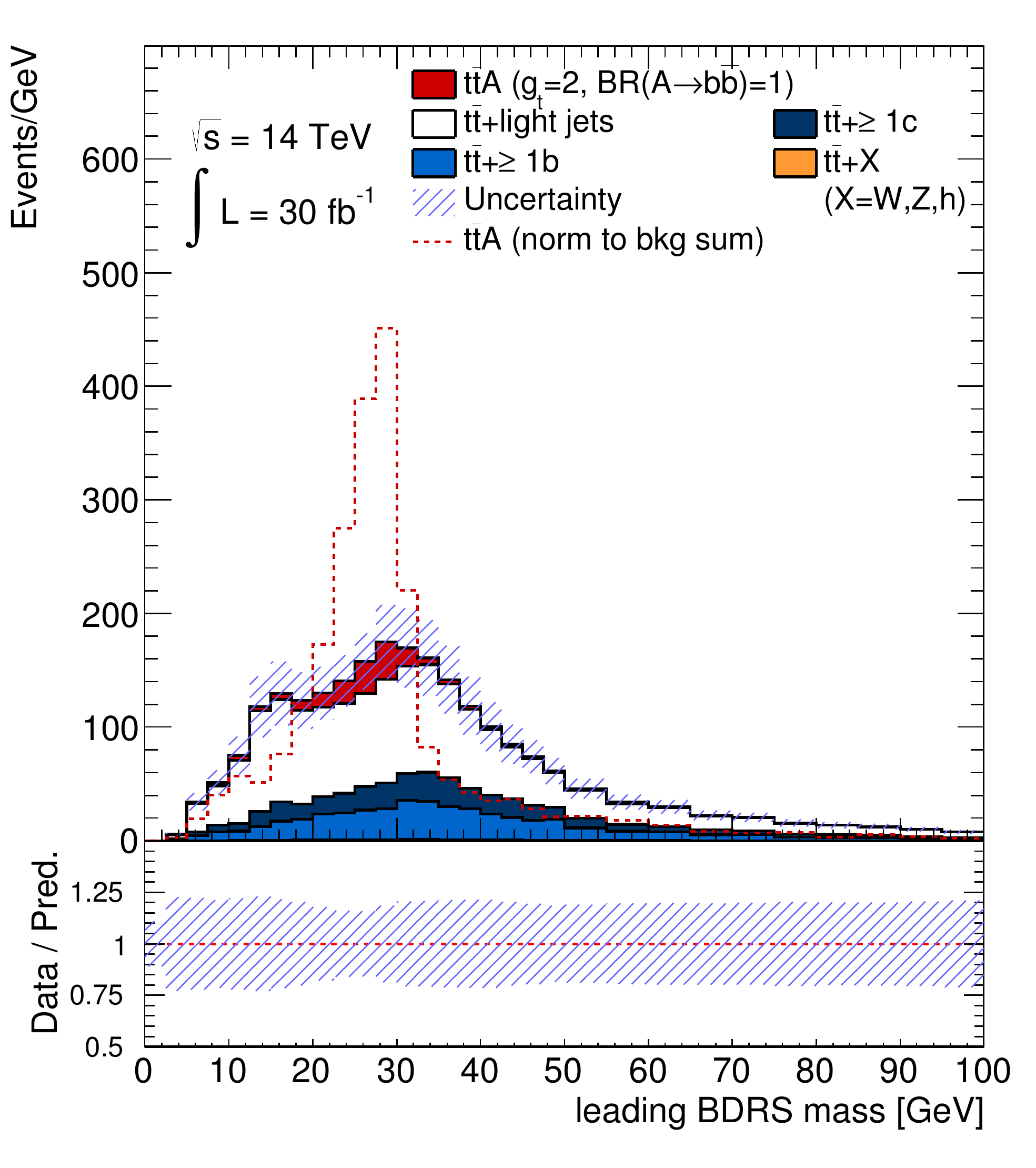} &
\includegraphics[width=0.3\textwidth]{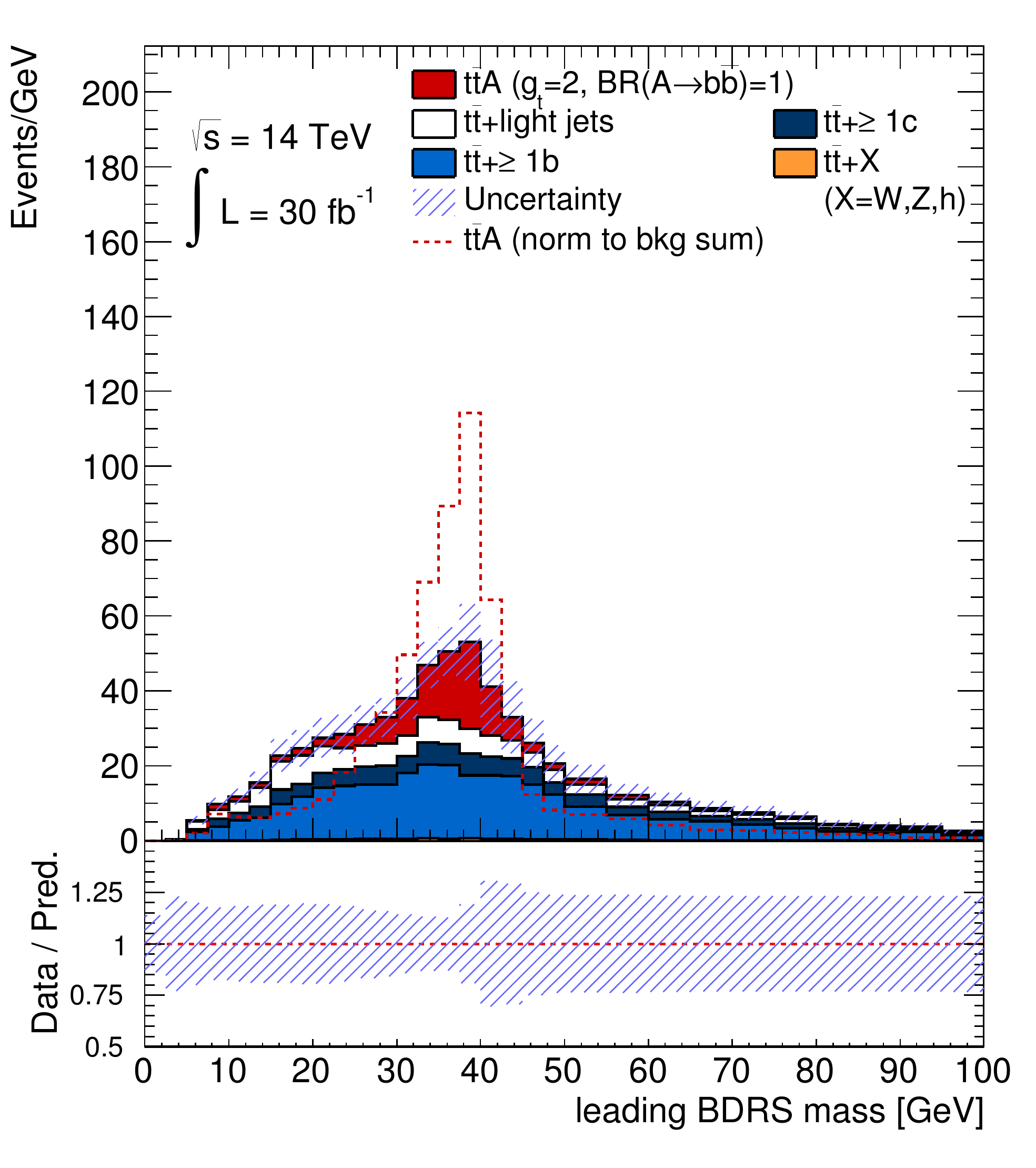} \\
\end{tabular}
\caption{\small {Distribution of the leading BDRS jet mass in the two analysis channels considered after final selection: 
(top) ($\geq$5j, 3b) and (bottom) ($\geq$5j, $\geq$4b), for different values of $m_A$ (20, 30 and 40 GeV).
The prediction corresponds to $\sqrt{s}=14$ TeV and an integrated luminosity of 30 fb$^{-1}$.
Several background categories have been merged for visibility. The expected contribution from 
the $\ttbar A$ signal under the assumptions of $g_t=2$ and ${\cal B}(A\to b\bar{b})=1$  is also shown
(red histogram), stacked on top of the SM background. The dashed red line shows the $\ttbar A$  signal 
distribution normalised to the background yield to better compare the shape to that of the background.
The bottom panel displays the expected total systematic uncertainty on the total prediction prior to the fit 
to the pseudo-data.}}
\label{fig:mA_1} 
\end{center}
\end{figure}

\begin{figure}[htbp]
\begin{center}
\begin{tabular}{ccc}
$m_A = 60$~GeV & $m_A = 80$~GeV &  $m_A = 100$~GeV \\
\includegraphics[width=0.3\textwidth]{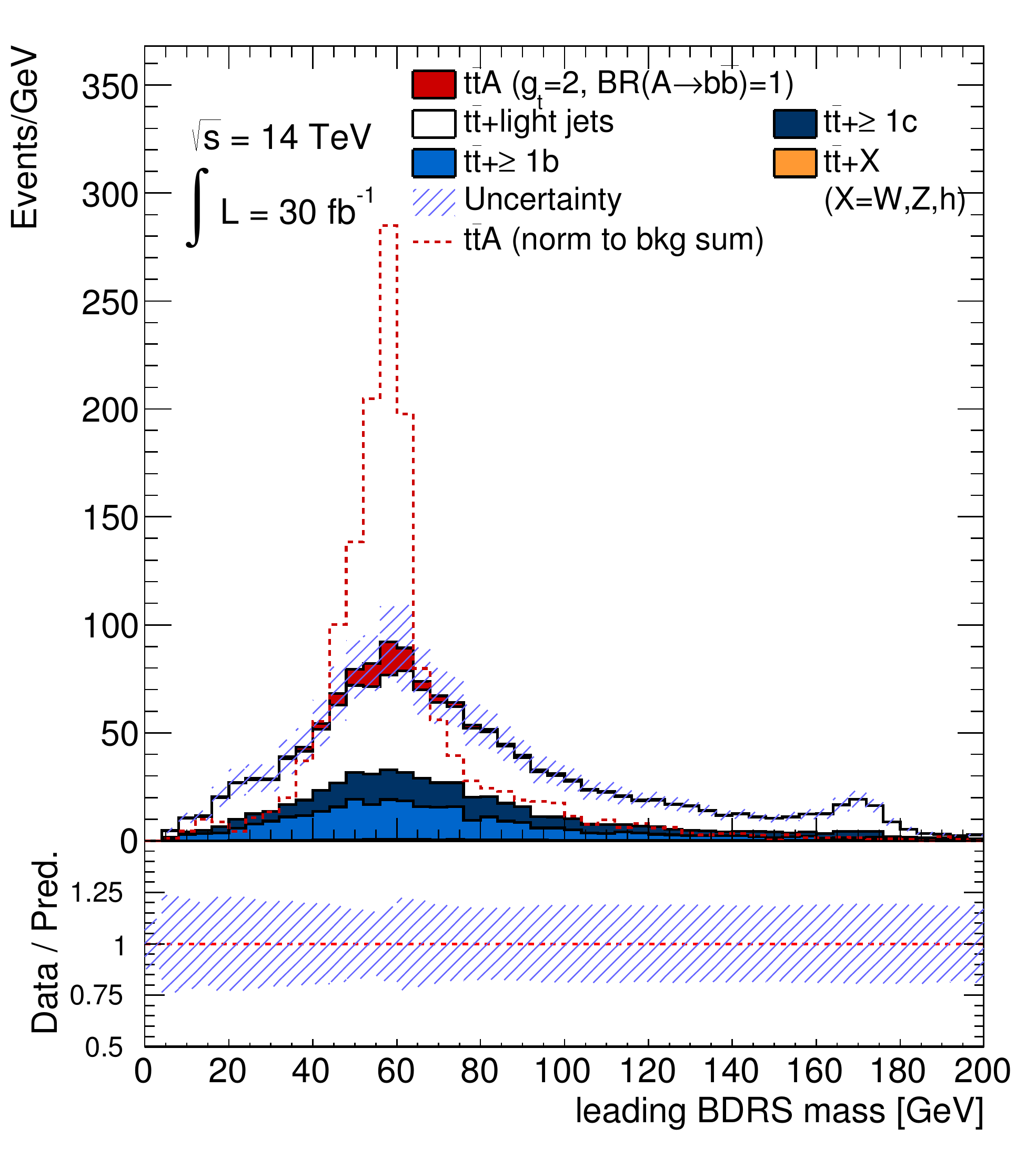} &
\includegraphics[width=0.3\textwidth]{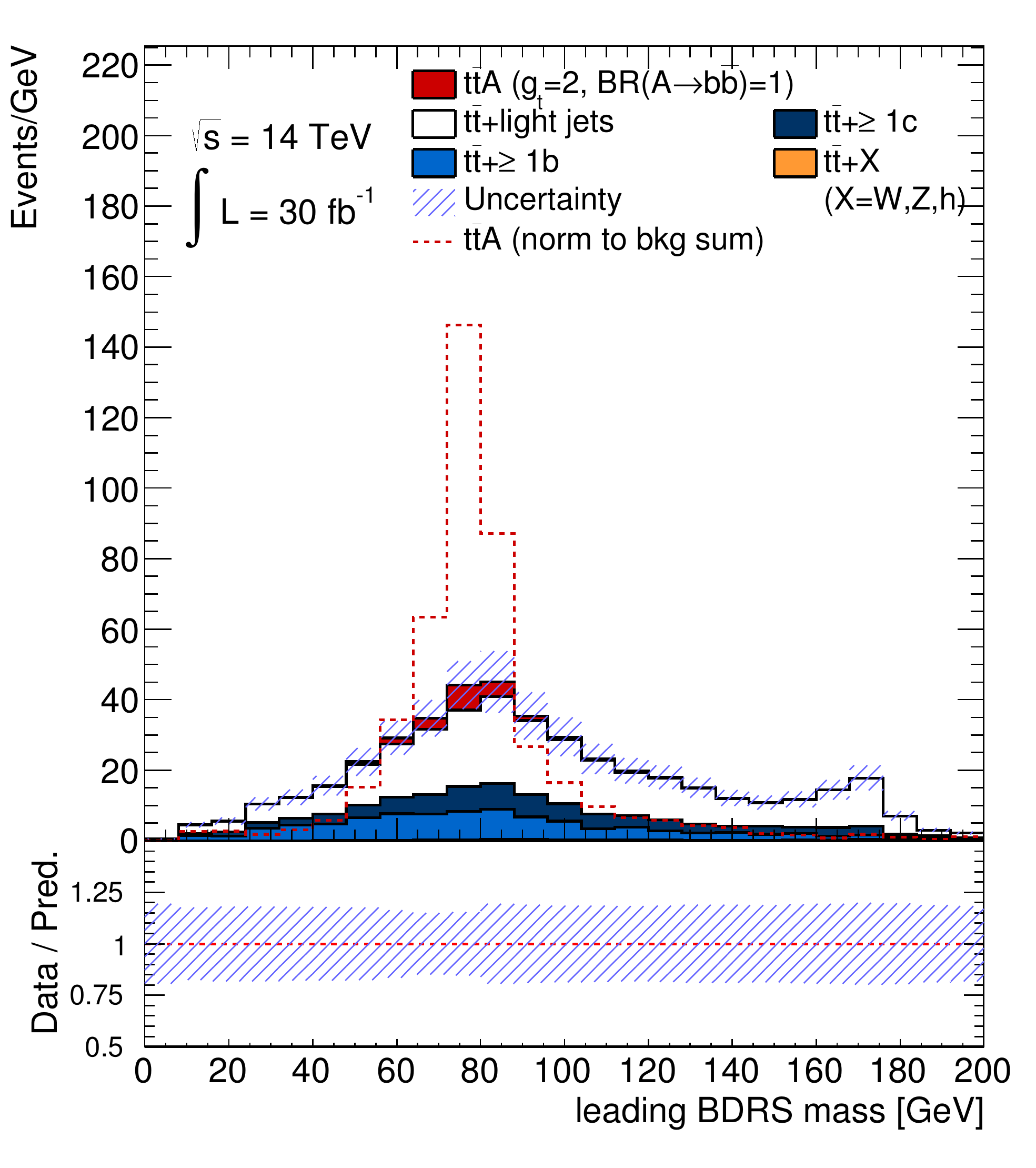} &
\includegraphics[width=0.3\textwidth]{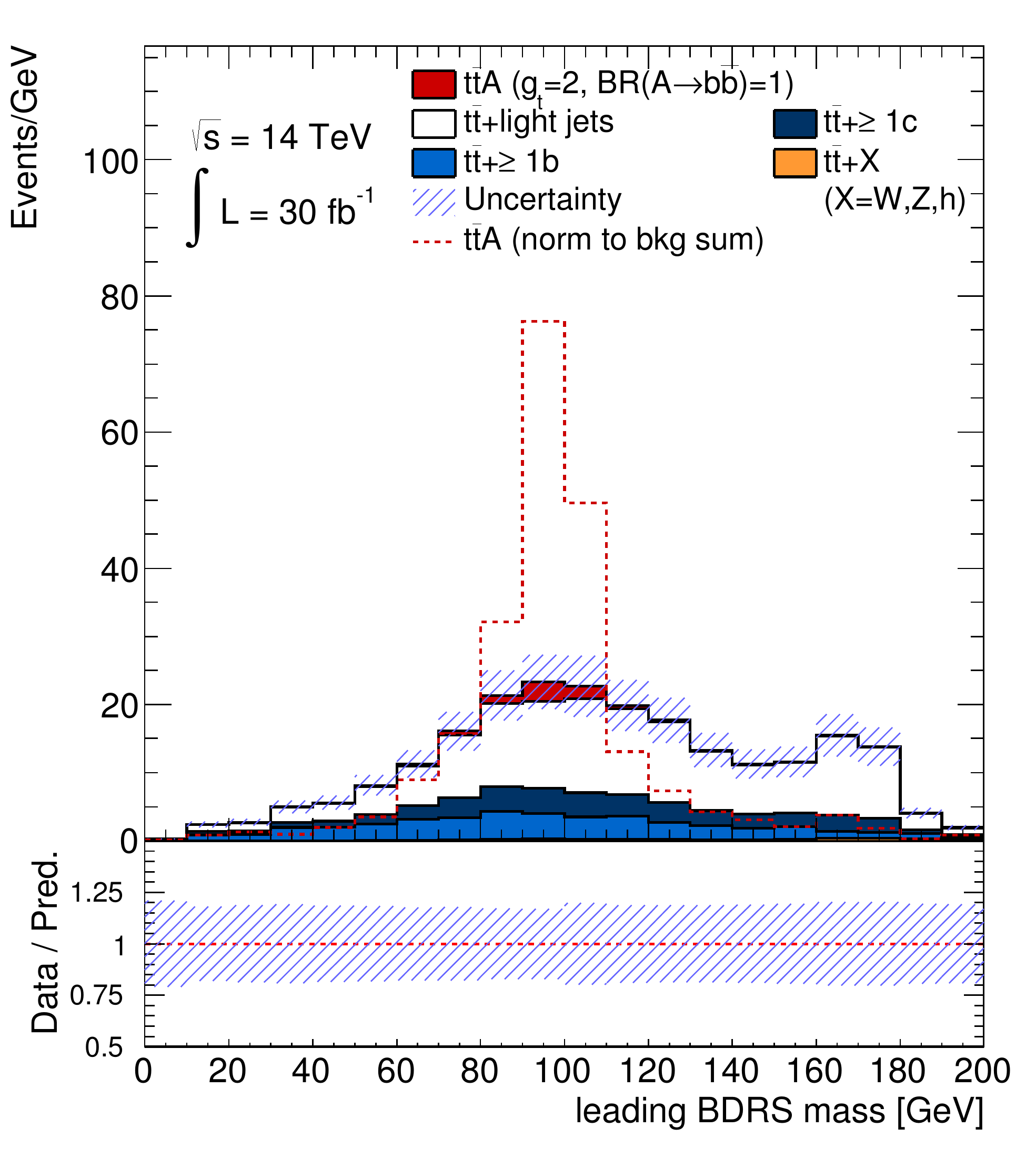} \\
\includegraphics[width=0.3\textwidth]{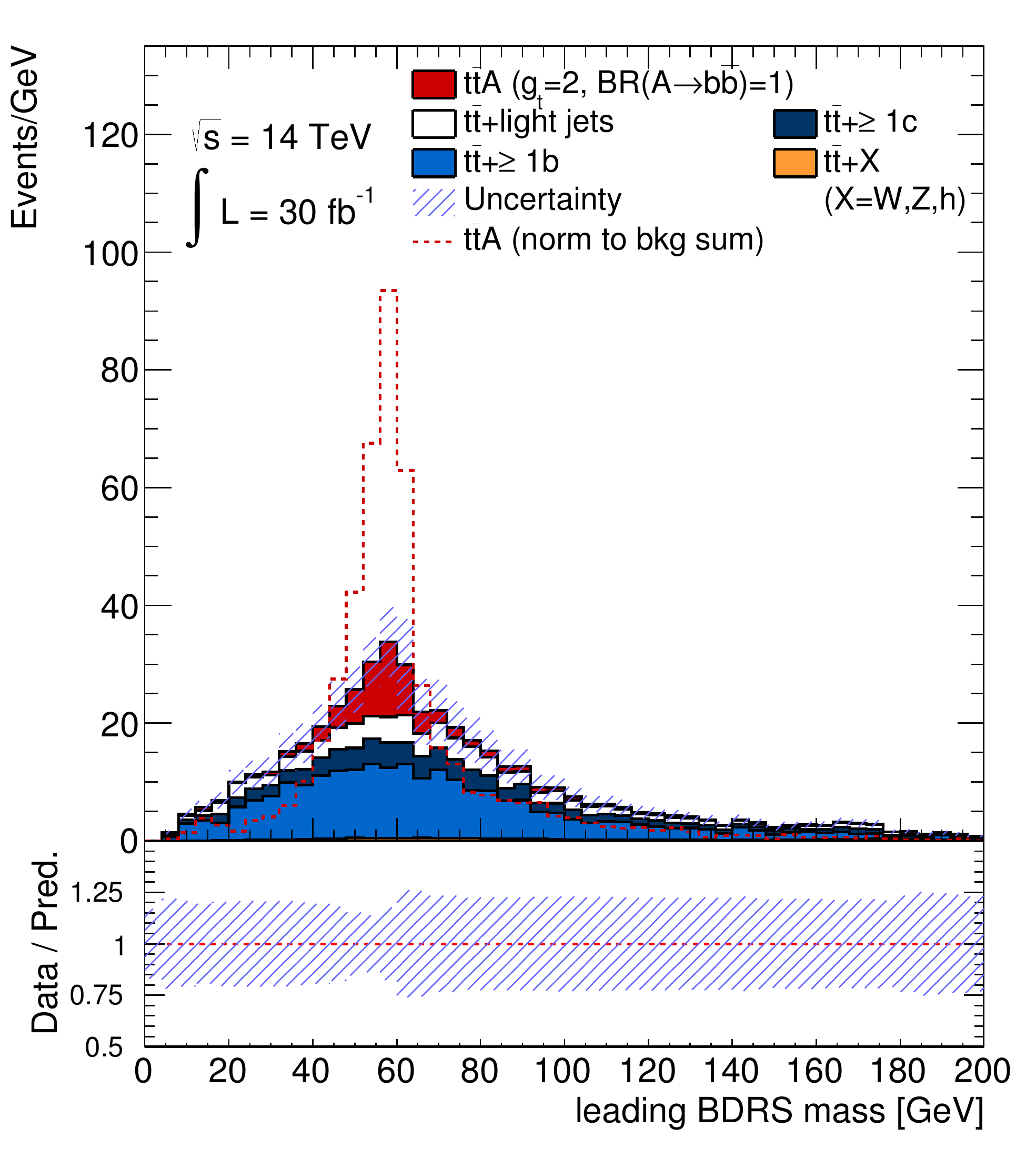} &
\includegraphics[width=0.3\textwidth]{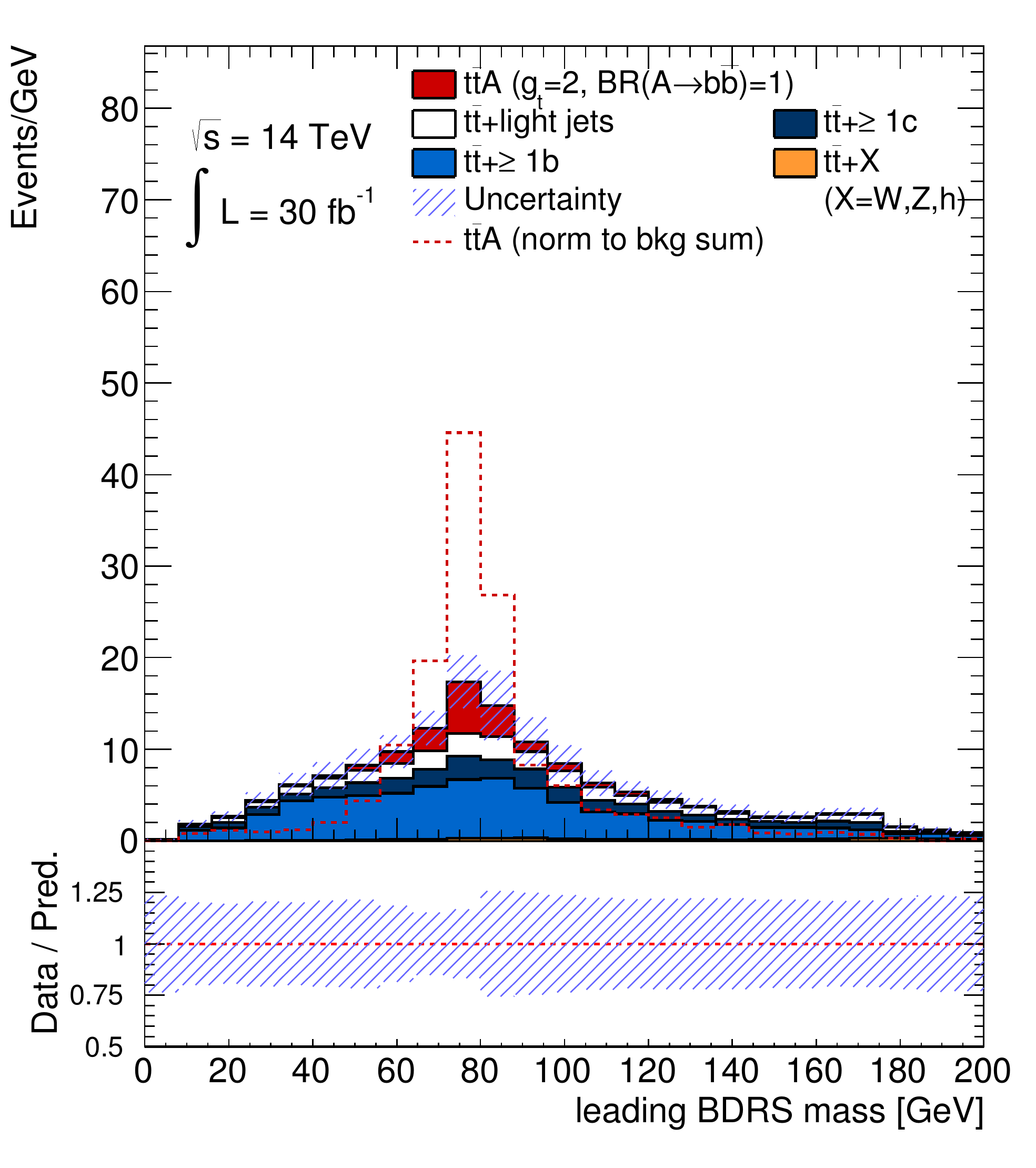} &
\includegraphics[width=0.3\textwidth]{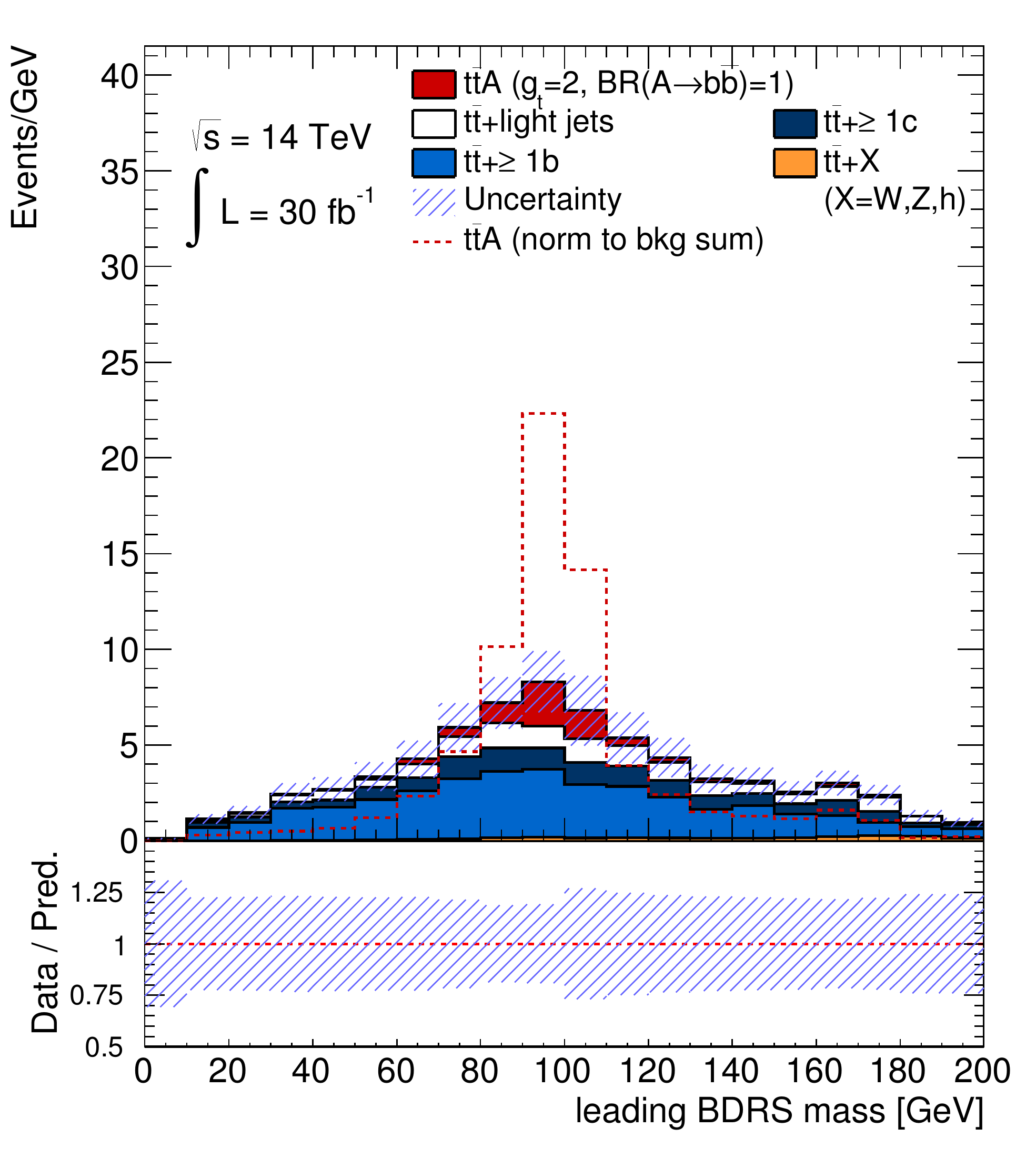} \\
\end{tabular}
\caption{\small {Distribution of the leading BDRS jet mass in the two analysis channels considered after final selection: 
(top) ($\geq$5j, 3b) and (bottom) ($\geq$5j, $\geq$4b), for different values of $m_A$ (60, 80 and 100 GeV).
The prediction corresponds to $\sqrt{s}=14$ TeV and an integrated luminosity of 30 fb$^{-1}$.
Several background categories have been merged for visibility. The expected contribution from 
the $\ttbar A$ signal under the assumptions of $g_t=2$ and ${\cal B}(A\to b\bar{b})=1$  is also shown
(red histogram), stacked on top of the SM background. The dashed red line shows the $\ttbar A$  signal 
distribution normalised to the background yield to better compare the shape to that of the background.
The bottom panel displays the expected total systematic uncertainty on the total prediction prior to the fit 
to the pseudo-data.}}
\label{fig:mA_2} 
\end{center}
\end{figure}

The final discriminating variable is the invariant mass of the selected C/A jet, referred to as 
``leading BDRS jet mass". Figures~\ref{fig:mA_1} and~\ref{fig:mA_2} show the expected distribution of the BDRS jet mass
for signal and background in each of the analysis channels, for the different $m_A$ values considered. The distributions
correspond to $\sqrt{s}=14$ TeV and are normalised to an integrated luminosity of 30 fb$^{-1}$.
For the assumed values of $g_t=2$ and ${\cal B}(A\to b\bar{b})=1$, the signal is clearly visible
on top of the background.

\begin{table}[h] 
\begin{center} 
\begin{tabular}{ccccc|cc} 
\hline\hline
&$\quad$$\ttbar$+$\geq$$1b$$\quad$ &$\quad$$\ttbar$+$\geq$$1c$$\quad$&$\quad$$\ttbar$+light-jets$\quad$&$\quad$$\ttbar+X$$\quad$ & $\quad$Total bkg.$\quad$ & $\quad$$\ttbar A$$\quad$ \\ 
\hline\hline
\multicolumn{7}{c}{$m_A=30$~GeV} \\
\hline
1 lepton&$4167$&$10958$&$155648$&$299$&$171072$&$377$ \\ 
$\geq$5 jets&$3109$&$7678$&$61866$&$215$&$72868$&$268$ \\ 
\hline
3 $b$-tags&$766$&$765$&$2702$&$30.1$&$4263$&$72.4$ \\ 
$\geq$1 CA6 jets &$510$&$502$&$1485$&$21.4$&$2518$&$55.7$ \\ 
$\geq$2 $b$-tags in selected CA6 jet & $45.1$&$38.4$&$159$&$1.9$&$\bf 245$&$\bf 14.6$ \\ 
\hline
$\geq$4 $b$-tags&$234$&$100$&$128$&$10.6$&$474$&$28.7$ \\ 
$\geq$1 CA6 jets&$171$&$70.1$&$75.7$&$7.9$&$325$&$23.8$ \\ 
$\geq$2 $b$-tags in selected CA6 jet &$36.9$&$13.2$&$18.5$&$1.5$&$\bf 70.2$&$\bf 11.7$ \\ 
\hline\hline
\multicolumn{7}{c}{$m_A=80$~GeV} \\
\hline
1 lepton&$4167$&$10958$&$155648$&$299$&$171072$&$240$ \\ 
$\geq$5 jets&$3109$&$7678$&$61866$&$215$&$72868$&$198$ \\ 
\hline
3 $b$-tags&$766$&$765$&$2702$&$30.1$&$4263$&$57.5$ \\ 
$\geq$1 CA8 jets &$252$&$246$&$646$&$11.5$&$1155$&$23.6$ \\ 
$\geq$2 $b$-tags in selected CA8 jet &$32.3$&$32.8$&$125$&$2.0$&$\bf 192$&$\bf 6.1$ \\ 
\hline
$\geq$4 $b$-tags&$234$&$100$&$128$&$10.6$&$474$&$25.0$ \\ 
$\geq$1 CA8 jets&$91.6$&$36.4$&$35.0$&$4.3$&$167$&$11.6$ \\ 
$\geq$2 $b$-tags in selected CA8 jet &$25.8$&$10.6$&$12.6$&$1.5$&$\bf 50.4$&$\bf 5.3$ \\ 
\hline\hline
\end{tabular} 
\caption{\small {Expected signal and SM backgrounds at $\sqrt{s}=14$ TeV
per fb$^{-1}$ of integrated luminosity as a function of the selection cuts applied in each
of the analysis channels under consideration (see text for details): ($\geq$5j, 3b) and ($\geq$5j, $\geq$4b).
The signal prediction is obtained under the assumptions of $g_t=2$ and ${\cal B}(A\to b\bar{b})=1$.
Several background categories have been merged for readability. The sum of 
$\ttbar$+$W$, $\ttbar$+$Z$ and $\ttbar$+$h_{\rm SM}$ is denoted as $\ttbar$+$X$. 
The yields shown correspond to the optimised selections for two different values
of $m_A$, 30 GeV and 80 GeV. Shown in bold are the signal and backgrounds
expectations after full selection in each of the analysis channels considered.}}
\label{tab:yields} 
\end{center} 
\end{table} 

\subsection{Systematic uncertainties}
\label{sec:systematics}

Several sources of systematic uncertainty are considered that can affect the normalisation of signal 
and background and/or the shape of the BDRS jet mass distribution. 
Individual sources of systematic uncertainty are considered uncorrelated.  Correlations of a given 
systematic uncertainty are maintained across processes and analysis channels. 
The choices of what uncertainties to consider and their magnitude are inspired by recent 
$\ttbar$+$h_{\rm SM}$, $h_{\rm SM}\to b\bar{b}$ searches at the LHC~\cite{Aad:2015gra}.

A 15\% normalisation uncertainty is assigned to $\ttbar$+light-jets corresponding to the modelling of the jet multiplicity 
spectrum. A 30\% normalisation uncertainty is assigned to each of the $\ttbar$+HF components ($\ttbar$+$b$, 
$\ttbar$+$b\bar{b}$, $\ttbar$+$B$, $\ttbar$+$c$, $\ttbar$+$c\bar{c}$, $\ttbar$+$C$), and taken to be uncorrelated among 
them. These uncertainties are expected to be conservative given the recent progress in NLO predictions for $\ttbar$ with
up to two jets merged with a parton shower~\cite{Hoeche:2014qda}, as well as NLO predictions for $\ttbar$+$\geq 1b$ production 
in the 4F scheme matched to a parton shower~\cite{Cascioli:2013era}. Cross section uncertainties for $\ttbar$+$W$, $\ttbar$+$Z$ and
$\ttbar$+$h_{\rm SM}$ are taken to be 30\% for each process. Uncertainties associated to jet energy and mass calibration are taken
to be 5\% per jet, fully correlated between energy and mass and across all jets in the event. Finally, uncertainties on the
$b$-, $c$- and light-jet tagging efficiencies are taken to be 3\%, 6\% and 15\% respectively. These uncertainties are taken 
as uncorrelated between $b$-jets, $c$-jets, and light-jets. 
As shown in Figs.~\ref{fig:mA_1} and~\ref{fig:mA_2}, the resulting total background normalisation uncertainty is about 20\%, 
although the different uncertainty components have different shape in the final distribution.

\subsection{Statistical analysis}
\label{sec:stat_analysis}

The BDRS jet mass distribution in the two analysis channels 
under consideration (see Figs.~\ref{fig:mA_1} and~\ref{fig:mA_2}) are tested for the presence of
a signal. To obtain the most realistic possible sensitivity projection, a sophisticated statistical analysis   
is performed, following very closely the strategy adopted in the experimental analyses at the LHC.

For each $m_A$ hypothesis, 95\% CL upper limits on the $\ttbar A$ production cross section times branching ratio, 
$\sigma(\ttbar A) \times {\cal B}(A \to b\bar{b})$, are obtained with the CL$_{\rm{s}}$ method~\cite{Junk:1999kv,Read:2002hq} 
using a profile likelihood ratio as test statistic implemented in the {\sc RooFit} package~\cite{Verkerke:2003ir,RooFitManual}. 
The likelihood function ${\cal L}(\mu,\theta)$ depends on 
the signal-strength parameter $\mu$,  a multiplicative factor to the theoretical signal production cross section,
and $\theta$, a set of nuisance parameters that encode the effect of systematic uncertainties in the analysis.
The likelihood function is constructed as a product of Poisson probability terms over all bins of the distributions analysed, 
and of Gaussian or log-normal probability terms, each corresponding to a nuisance parameter. 
For a given assumed value of $\mu$, the profile likelihood ratio $q_\mu$ is defined as:
\begin{equation}
q_\mu = -2\ln({\cal L}(\mu,\hat{\hat{\theta}}_\mu)/{\cal L}(\hat{\mu},\hat{\theta})), 
\end{equation}
where $\hat{\hat{\theta}}_\mu$ are the values of the nuisance parameters that maximise the likelihood 
function for a given value of $\mu$, and $\hat{\mu}$ and $\hat{\theta}$ are the values of the parameters 
that maximise the likelihood function (with the constraint $0\leq \hat{\mu} \leq \mu$).
The maximisation of the likelihood function over the nuisance parameters allows variations of the expectations 
for signal and background in order to improve the agreement with (pseudo-)data, yielding a
background prediction with reduced overall uncertainty and thus resulting in an improved sensitivity.
For a given $m_A$ hypothesis, values of the production cross section (parameterised by $\mu$) yielding 
CL$_{\rm{s}}$$<$0.05, where CL$_{\rm{s}}$ is computed using the asymptotic 
approximation~\cite{Cowan:2010js}, are excluded at $\geq$95\% CL.

\section{Estimated limits on a light CP-odd scalar}
\label{sec:limit}

\begin{figure}[htbp]
\begin{center}
\begin{tabular}{cc}
\includegraphics[width=0.45\textwidth]{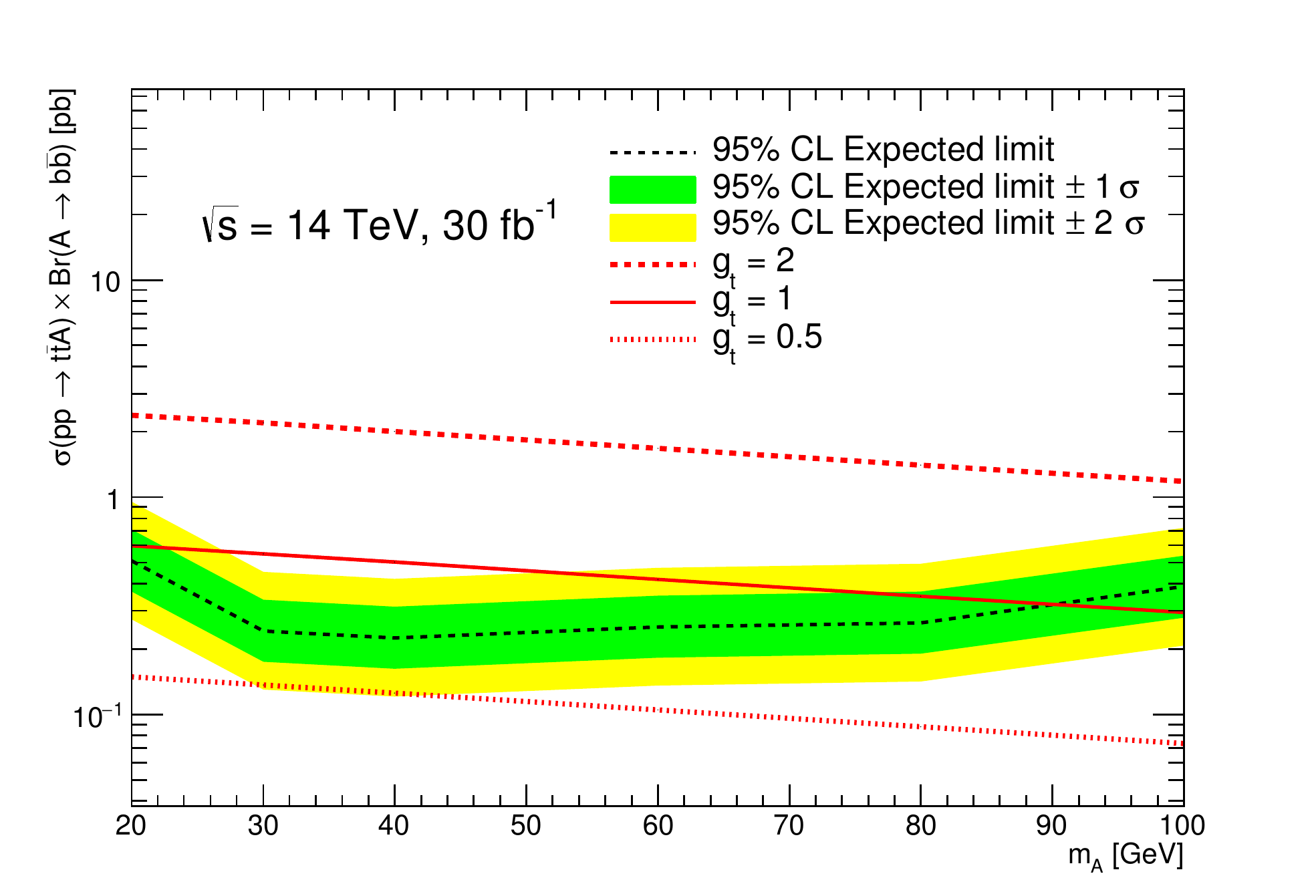} &
\includegraphics[width=0.45\textwidth]{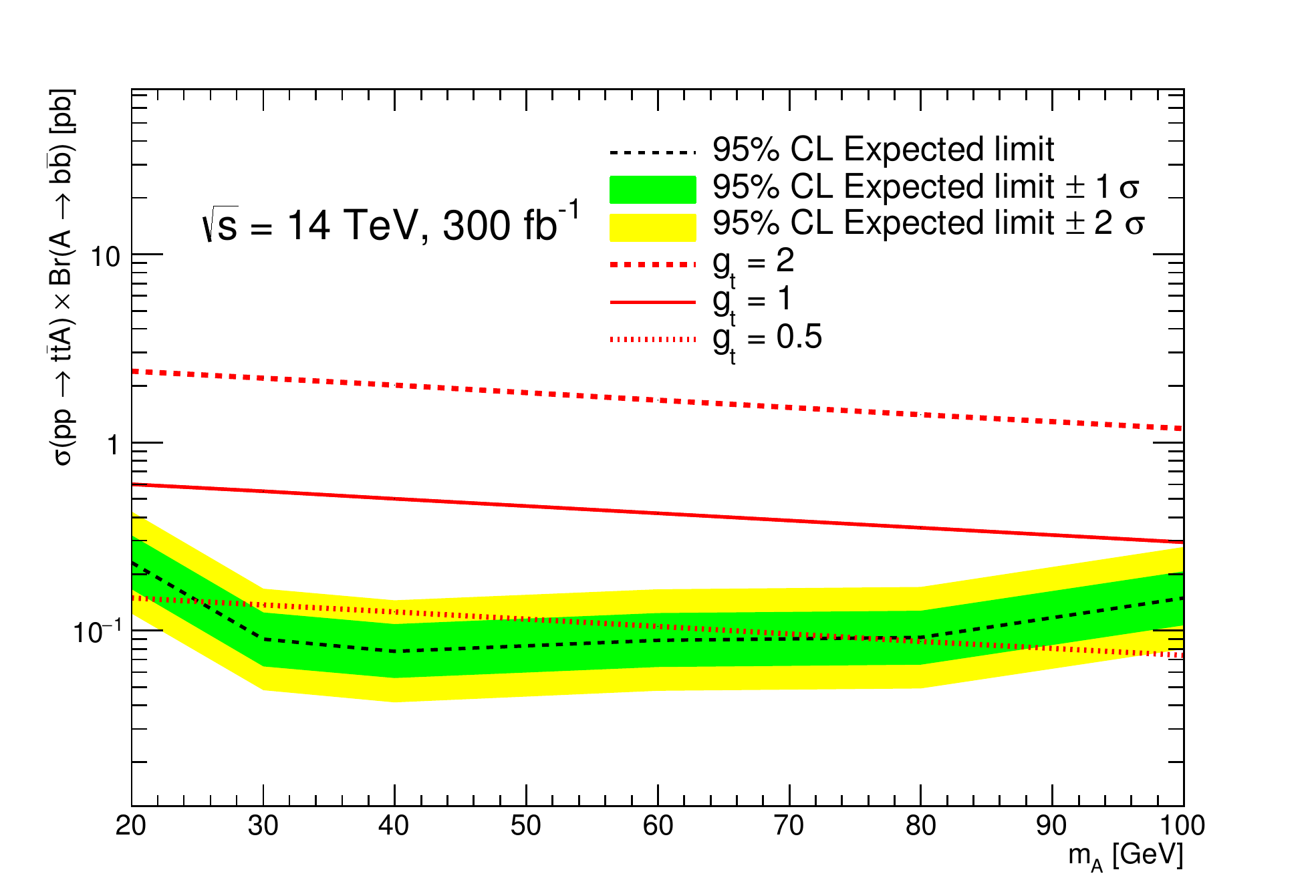} \\
(a) & (b) \\
\end{tabular}
\caption{\small {Expected 95\% CL upper limits on $\sigma(\ttbar A) \times {\cal B}(A\to b\bar{b})$ as a function of $m_A$ 
in $pp$ collisions at $\sqrt{s}=14$ TeV for an integrated luminosity of (a) 30 fb$^{-1}$ and (b) 300 fb$^{-1}$. 
The green and yellow bands correspond to 1 and 2 standard deviations respectively around the median expected limit 
under the background-only hypothesis. Also shown are the theoretical cross sections for $\sigma(\ttbar A)$ for different  
assumed values of $g_t$ (0.5, 1.0 and 2.0) and ${\cal B}(A\to b\bar{b})=1$.}}
\label{fig:limit_plots}
\end{center}
\end{figure}

Following the analyses steps and the limit setting outlined in Sects.~\ref{sec:model}-\ref{sec:analysis}, we estimate expected 95\% CL upper limits
on the production cross section times branching ratio, $\sigma(\ttbar A) \times {\cal B}(A\to b\bar{b})$, as a function of $m_A$ (see Fig.~\ref{fig:limit_plots}).
Table~\ref{tab:sigma_95CL} summarises the 95\% CL upper limits on $\sigma(\ttbar A) \times {\cal B}(A\to b\bar{b})$ as a function of $m_A$ for different values
of the integrated luminosity. Under the assumption ${\cal B}(A\to b\bar{b})=1$, the upper limits on $\sigma(\ttbar A) \times {\cal B}(A\to b\bar{b})$ 
can be translated on upper limits on $|g_t|$, which are summarised in Table~\ref{tab:ct_95CL}.

Using the reconstruction strategy outlined in Sec.~\ref{subsec:analysis}, a CP-odd scalar that couples with $g_t=1$ can be excluded for $20 \leq m_A \leq 90$ GeV with only $30~\mathrm{fb}^{-1}$ of data (see Fig.~\ref{fig:limit_plots}). With an increased statistics of $300~\mathrm{fb}^{-1}$ couplings as low as $g_t \simeq 0.5$ can be constrained over a large mass range, i.e. $30 \leq m_A \leq 80$ GeV.

\begin{table}[h] 
\begin{center} 
\begin{tabular}{ccccccc} 
\hline\hline
 & \multicolumn{6}{c}{95\% CL upper limits on $\sigma(\ttbar A) \times {\cal B}(A\to b\bar{b})$ (pb)} \\
\hline
  & \multicolumn{6}{c}{$m_A$ (GeV)} \\
  \cline{2-7} 
${\cal L}$ (fb$^{-1}$) & $\quad$ 20 $\quad$ & $\quad$ 30 $\quad$ & $\quad$ 40 $\quad$ & $\quad$ 60 $\quad$ & $\quad$ 80 $\quad$ & $\quad$ 100 $\quad$ \\
\hline
1 & 4.46 & 2.50 & 2.38 & 2.57 & 2.78 & 3.94 \\
30 & 1.02 & 0.48 & 0.45 & 0.51 & 0.53 & 0.78  \\
100 & 0.67 & 0.29 & 0.25 & 0.29 & 0.30 & 0.46 \\
300 & 0.46 & 0.18 & 0.16 & 0.18 & 0.18 & 0.30 \\
3000 & 0.17 & 0.066 & 0.057 & 0.065 & 0.065 & 0.13 \\
\hline\hline
\end{tabular} 
\caption{\small {Expected 95\% CL upper limits on $\sigma(\ttbar A) \times {\cal B}(A\to b\bar{b})$ as a function of $m_A$ 
in $pp$ collisions at $\sqrt{s}=14$ TeV for different integrated luminosities.}}
\label{tab:sigma_95CL} 
\end{center} 
\end{table} 

\begin{table}[h!] 
\begin{center} 
\begin{tabular}{ccccccc} 
\hline\hline
 & \multicolumn{6}{c}{95\% CL upper limits on $|g_t|$} \\
\hline
  & \multicolumn{6}{c}{$m_A$ (GeV)} \\
  \cline{2-7} 
${\cal L}$ (fb$^{-1}$) & $\quad$ 20 $\quad$ & $\quad$ 30 $\quad$ & $\quad$ 40 $\quad$ & $\quad$ 60 $\quad$ & $\quad$ 80 $\quad$ & $\quad$ 100 $\quad$ \\
\hline
1 & 2.73  & 2.14 & 2.18 & 2.48 & 2.82 & 3.65 \\
30 & 1.31  & 0.94 & 0.95 & 1.10 & 1.23 & 1.62 \\
100 & 1.06 & 0.72 & 0.71 & 0.83 & 0.93 & 1.25 \\
300 & 0.88 & 0.57 & 0.55 & 0.65 & 0.72 & 1.00 \\
3000 & 0.54 & 0.35 & 0.34 & 0.39 & 0.43 & 0.67 \\
\hline\hline
\end{tabular} 
\caption{\small {Expected 95\% CL upper limits on $|g_t|$ as a function of $m_A$ 
in $pp$ collisions at $\sqrt{s}=14$ TeV for different integrated luminosities, under the assumption ${\cal B}(A\to b\bar{b})=1$.}}
\label{tab:ct_95CL} 
\end{center} 
\end{table} 

\section{Interpretation of limits}
\label{sec:interpretation}

A light CP-odd Higgs boson ($m_A < 125$~GeV), which may or may not be related to global symmetries being present, exists in many extensions of the SM. Its couplings with gauge bosons are generically suppressed, yielding weak bounds from LEP. If $m_A<m_{h_{\rm SM}}/2$, it may be searched via the 
decay $h_{\rm SM} \to AA$. Though such decay sometimes has a large branching ratio, being in conflict with current Higgs precision data, there do exist scenarios, in both supersymmetric and non-supersymmetric theories, where the ${\cal B}(h_{\rm SM}\to AA)$ is suppressed. 
Therefore, new strategies for collider searches that could cover as large as possible model parameter space with a light CP-odd Higgs boson, are necessary. 
Next, we will interpret our collider analysis of $t\bar t A$ in several representative beyond-SM scenarios. 

\subsection{2HDM}

In the MSSM, a supersymmetric extension of a type-II 2HDM, a scenario with a light CP-odd Higgs boson is hard to achieve, given constraints from precision Higgs data. This is not surprising since there are only two free parameters at tree level in the Higgs sector, due to supersymmetric interrelations. The picture, however, is changed in the 2HDM without supersymmetry. With a softly-broken $Z_2$ symmetry ($\Phi_1 \to \Phi_1$, $\Phi_2 \to -\Phi_2$), which is often introduced to suppress scalar-mediated flavor changing processes, the Higgs potential of the 2HDM is given by: 
\begin{eqnarray} 
V(\Phi_1,\Phi_2) &=& m^2_1 \Phi^{\dagger}_1\Phi_1+m^2_2
\Phi^{\dagger}_2\Phi_2 + (m^2_{12} \Phi^{\dagger}_1\Phi_2+{\mathrm{h.c.}
}) +\frac{1}{2} \lambda_1 (\Phi^{\dagger}_1\Phi_1)^2 +\frac{1}{2}
\lambda_2 (\Phi^{\dagger}_2\Phi_2)^2\nonumber \\ 
&& +\lambda_3
(\Phi^{\dagger}_1\Phi_1)(\Phi^{\dagger}_2\Phi_2) + \lambda_4
(\Phi^{\dagger}_1\Phi_2)(\Phi^{\dagger}_2\Phi_1) + \frac{1}{2}
\lambda_5[(\Phi^{\dagger}_1\Phi_2)^2+{\mathrm{h.c.}}].
\end{eqnarray}
Here $\Phi_{1,2}$ are complex $SU(2)_L$ doublets. Assuming no CP-violation, the model has two CP-even and one CP-odd spin-0 neutral eigenstates, denoted as $h$, $H$, and $A$, respectively. Such a setup contains seven free parameters at tree level (including all Higgs masses), yielding a large parameter space that can accommodate a light CP-odd Higgs boson. 

Theoretically, the 125 GeV SM-like Higgs boson $h_{\rm SM}$ could be either the light CP-even Higgs boson ($h$) or the heavy one ($H$). If $m_A < m_{h_{\rm SM}}/2$, the decay $h_{\rm SM}\to AA$ is kinematically allowed. Often the partial width for $h_{\rm SM}\to AA$ becomes comparable or ever dominant over that of $h_{\rm SM}\to b\bar{b}$, given that the latter is suppressed by the lightness of the bottom quark.  Therefore, $h_{\rm SM} \to AA$ decays become a good probe for these light bosonic particles. However, as discussed recently~\cite{Bernon:2014nxa},\footnote{See Refs.~\cite{Draper:2010ew,Huang:2014cla} for discussions in the context of the NMSSM.} in the alignment limit [$\cos(\beta -\alpha)=0$ if $h_{\rm SM}=h$, and $\sin(\beta -\alpha)=0$ if $h_{\rm SM}=H$], which is favoured by current precision Higgs measurements, the Higgs coupling $g_{h_{\rm SM} AA}$ is reduced to: 
\begin{eqnarray}
\left|g_{h_{\rm SM} AA}\right | = \left |-\frac{2 m_A^2 + m_{h_{\rm SM}}^2  - 4 m_{12}^2/\sin 2\beta }{v}\right |.
\end{eqnarray}
In case that  $2 m_A^2 + m_{h_{\rm SM}}^2  \sim 4 m_{12}^2/\sin 2\beta$, the decay $h_{\rm SM}\to AA$ would be greatly suppressed. Therefore, collider strategies  are needed to probe these scenarios with $m_A < m_{\rm SM}/2$, as well as the scenarios with $m_A > m_{\rm SM}/2$.

\begin{figure}[htbp]
\begin{center}
\begin{tabular}{c}
\includegraphics[width=0.5\textwidth,natwidth=610,natheight=642]{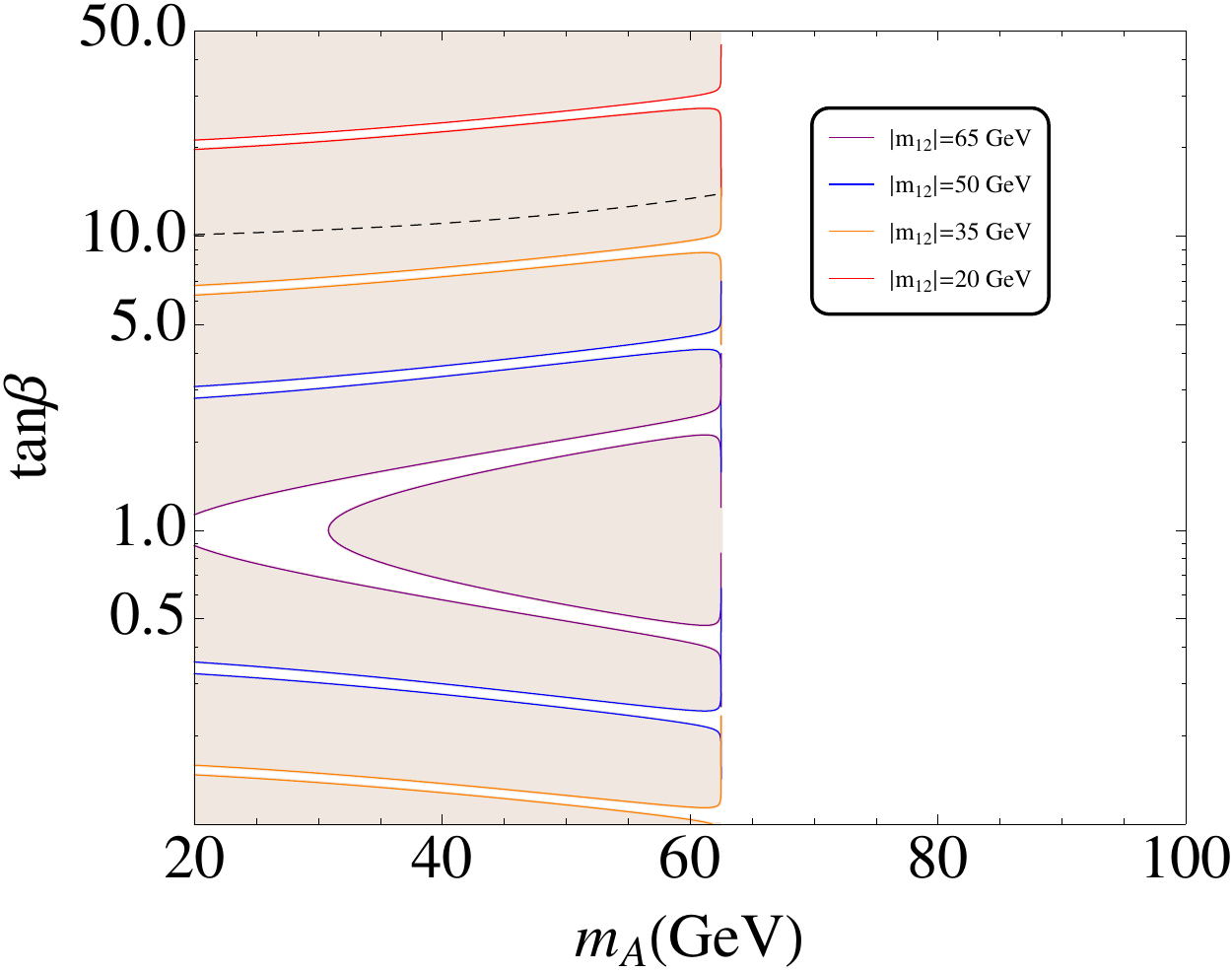} 
\end{tabular}
\caption{Parameter region with ${\cal B}(h_{\rm SM}\to AA)<30\%$ in the 2HDM (blank region). The blank belts in the region with $m_A < m_{h_{\rm SM}}/2$, which are characterised by different boundary colours, are yielded by different $m_{12}^2$ values. The black dashed line represents a universal upper limit on $\tan\beta$ due to the perturbation requirement for $\lambda_1$.}
\label{fig:Brhaa_2HDM} 
\end{center}
\end{figure}

We should note that the perturbation requirement for Higgs couplings yields bounds on $\tan\beta$. Particularly, the coupling $\lambda_1$ is related to the Higgs boson mass via the relation \cite{Bernon:2014nxa}: 
\begin{eqnarray}
\lambda_1 = \frac{m_h^2 + m_H^2 \tan^2\beta -  m_{12}^2 (\tan\beta +\tan^3 \beta)}{v^2}.
\end{eqnarray}
Assuming $g_{h_{\rm SM}\to AA}=0$, it becomes: 
\begin{eqnarray}
\lambda_1 = \frac{m_h^2 + \tan^2\beta (m_H^2 - m_{h_{\rm SM}}^2/2 - m_A^2)  }{v^2}.
\end{eqnarray}
Given $m_H^2 - m_{h_{\rm SM}}^2/2 - m_A^2 > 0$ for $m_A < m_{h_{\rm SM}}/2$, the perturbativity condition $\lambda_1 < 4\pi$ immediately sets an upper bound for $\tan\beta$ in this region:
\begin{eqnarray}
\tan\beta < \sqrt {\frac{4 \pi v^2 - m_h^2}{m_H^2 - m_{h_{\rm SM}}^2/2 - m_A^2 }} <  \sqrt {\frac{4 \pi v^2 }{m_{h_{\rm SM}}^2/2 - m_A^2 }}  \sim 10-20.
\end{eqnarray} 
These features are illustrated in Fig.~\ref{fig:Brhaa_2HDM}.  Additionally, the perturbation requirement for top Yukawa couplings can bound the $\tan\beta$ value from below. So we will limit our discussions for $\tan\beta > 0.1$. 

\begin{figure}[htbp]
\begin{center}
\begin{tabular}{cc}
\includegraphics[width=0.45\textwidth,natwidth=610,natheight=642]{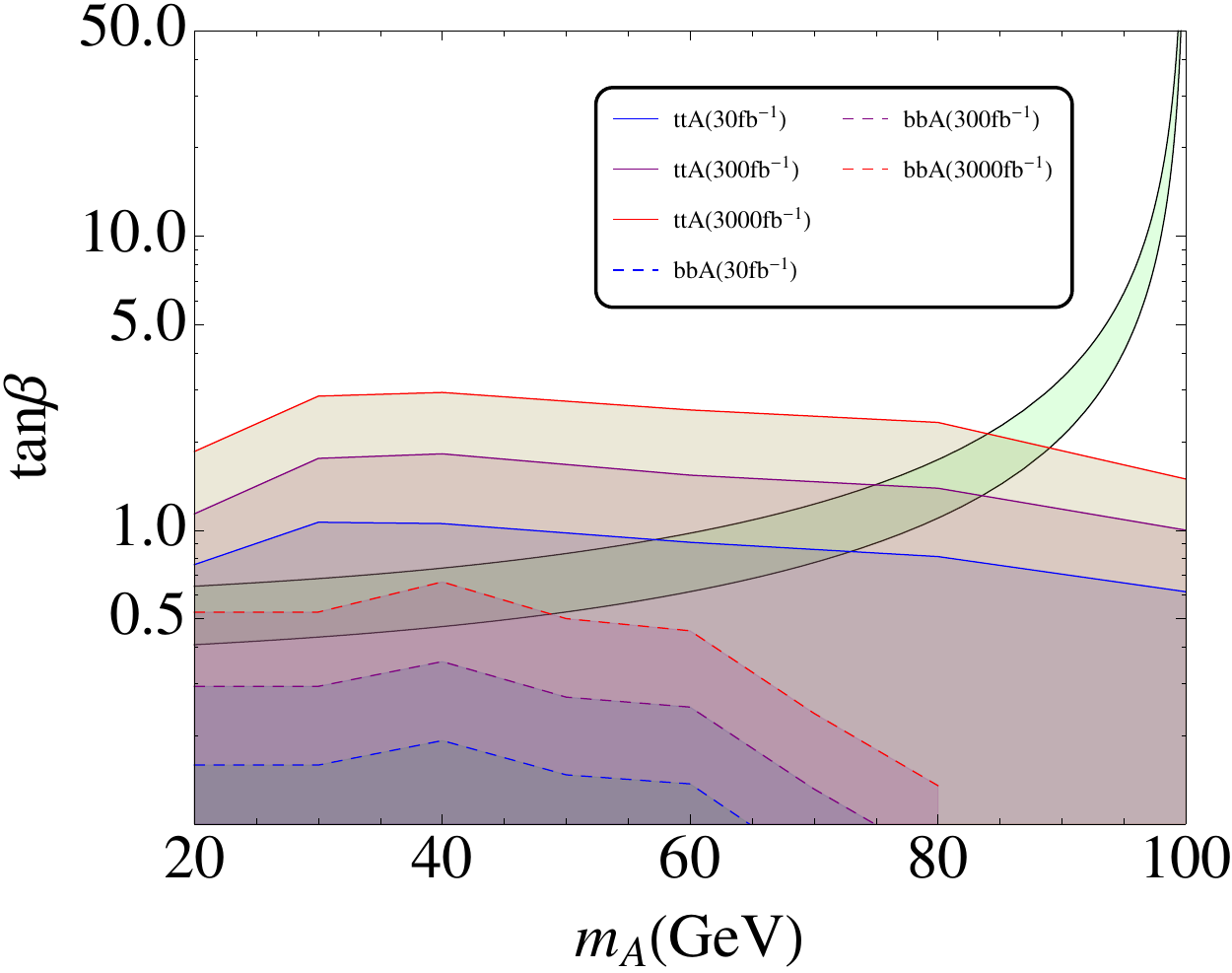} &
\includegraphics[width=0.45\textwidth,natwidth=610,natheight=642]{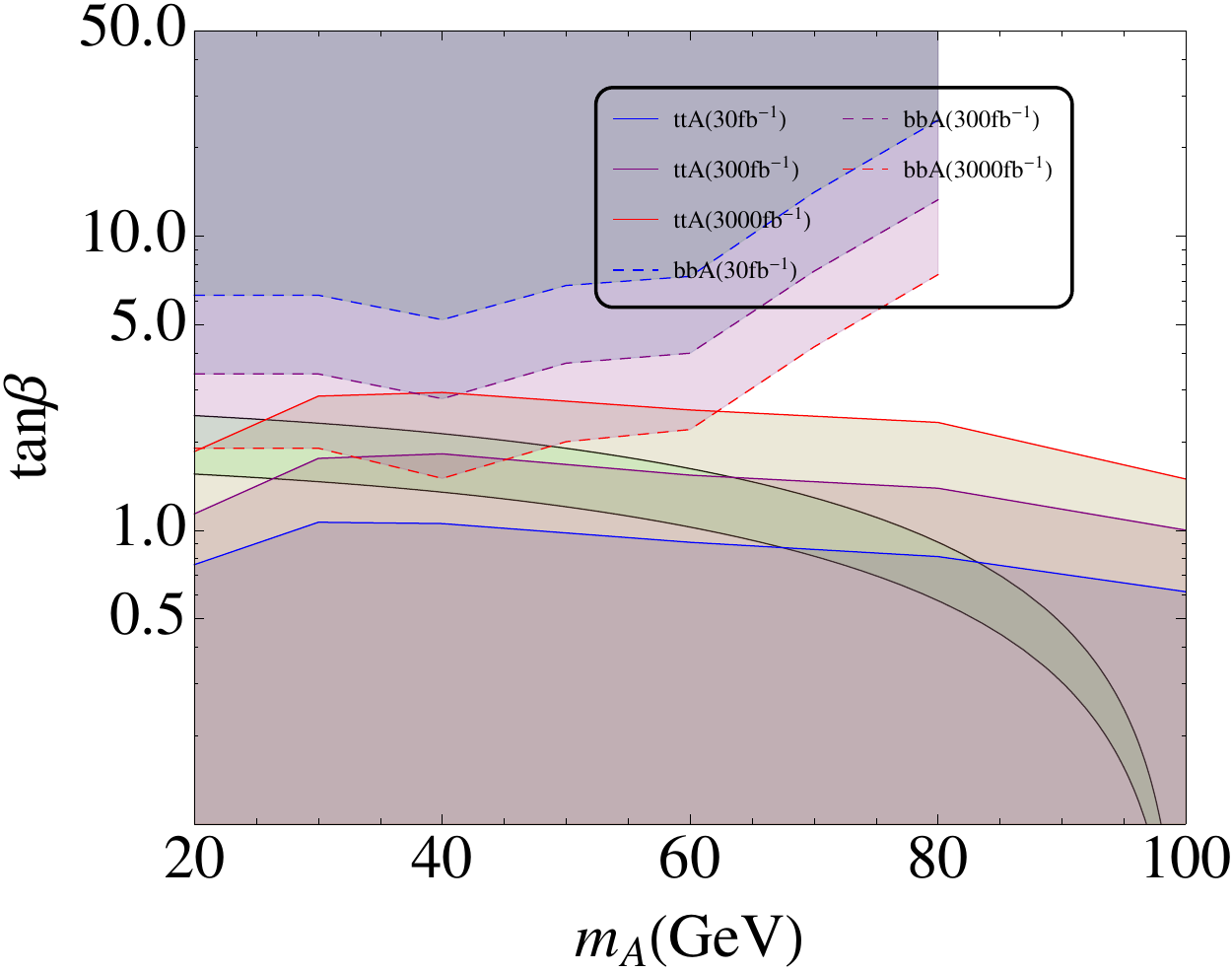} \\
(a) & (b) \\
\end{tabular}
\caption{Sensitivity reach (at 95\% CL) of the $b\bar {b} A$ and $t\bar{t}A$ channels within the (a) type-I 2HDM and (b) type-II 2HDM. The green bands represent a region where the recently observed gamma-ray excess from the Galactic Centre can be explained, yielding a DM annihilation cross section of $\langle \sigma v \rangle  \simeq 1-2.5 \times 10^{-26} {\rm cm}^3 {\rm s}^{-1}$. Here the DM particle mass $m_{\chi}=50$~GeV and the coupling between the mediator and the DM particles $y_\chi= 0.3$ are assumed.  }
\label{fig:reach_2HDM} 
\end{center}
\end{figure}

The expected sensitivities for probing these scenarios in the 2HDM via $b\bar bA$ and $t\bar{t}A$ production are presented in Fig.~\ref{fig:reach_2HDM}. For illustration, we focus on type-I and type-II 2HDMs. The $b\bar{b}A$ reach is estimated based on the projections from Ref.~\cite{Kozaczuk:2015bea}, neglecting
systematic uncertainties. Within a type-II  2HDM, the $t\bar{t}A$ and $b\bar{b}A$ channels are complementary to each other in searching for light CP-odd Higgs bosons, since the coupling $g_{bbA}$ is $\tan\beta$-enhanced whereas $g_{ttA}$ is $\cot \beta$-enhanced. With integrated luminosities in excess of 300~fb$^{-1}$, the whole parameter region can be covered except a corner with relatively large $m_A$ and moderate $\tan\beta$. This is interesting given that low $\tan\beta$ is particularly favoured by perturbativity. In contrast, within a type-I 2HDM, the coupling $g_{bbA}$ would also be $\cot\beta$-enhanced, so both search channels are no longer probing complementary $\tan\beta$ regions. As a matter of fact, in such scenario the $t\bar{t}A$ channel provides a better sensitivity to search for the light CP-odd Higgs boson over the whole mass range of $20~{\rm GeV} < m_A < 100~{\rm GeV}$, although the high-$\tan\beta$ region remains difficult to probe. 

Searches for $t\bar {t}A$ and $b\bar {b} A$ also provide a probe for DM physics. For example, consider a Dirac fermion $\chi$ that is a DM candidate, with mass $m_\chi$, and coupling to the CP-odd scalar $A$ via:
\begin{eqnarray}
\mathcal {L} \supset y_\chi A \bar \chi i  \gamma^5 \chi.
\end{eqnarray}
Integrating out $A$ yields a dimension-six effective operator:
\begin{eqnarray}
\mathcal {L}_{\rm eff} \sim  \frac{- y_b y_\chi m_b   }{\Lambda^3} \bar \chi   \gamma^5 \chi \bar b   \gamma^5 b.
\end{eqnarray}
Such an operator implies s-wave DM annihilation $\chi \chi \to b\bar{b}$ with
\begin{equation}
\left < \sigma v \right > = \frac{3}{8 \pi} \frac{y^2_\chi g_b^2 y_b^2 m_{\chi}^2}{(m_A^2 - 4 m_{\chi}^2)^2 + m_A^2 \Gamma_A^2} \sqrt{1-\frac{m_b^2}{m_\chi^2}},
\label{eq:anhil}
\end{equation}
allowing an explanation for the recently observed diffuse gamma-ray excess from the Galactic Centre \cite{Goodenough:2009gk,Vitale:2009hr},  and a spin-dependent and p-wave-suppressed direct detection signal, resulting in a weak bound from current direct detection searches. In Fig.~\ref{fig:reach_2HDM}, the $\tan\beta$--$m_A$ values consistent with an explanation of the gamma-ray excess are indicated, yielding a DM annhilation cross section of $\langle \sigma v \rangle  \simeq 1-2.5 \times 10^{-26} {\rm cm}^3 {\rm} s^{-1}$, with $m_{\chi}=50$~GeV \cite{Calore:2014nla} and  $y_\chi= 0.3$ assumed.  
In this scenario, monojet searches at the LHC would also be insensitive since the decay $A \to \chi \chi$ would be kinematically forbidden, while the $t\bar {t}A$, $A \to b\bar{b}$ search would provide an effective probe.

\subsection{NMSSM}

Another class of benchmark scenairos for light CP-odd Higgs bosons arise in the NMSSM, with the superpotential and soft
supersymmetry-breaking terms of its Higgs sector given by
\begin{align}
\mathbf{W} &= \lambda \mathbf{S} \mathbf{H_u} \mathbf{H_d} + \frac{1}{3}\kappa {\bf S}^3, 
\nonumber \\
V_{\text{soft}} &= {m^2_{H_d}} |H_d|^2 + {m^2_{H_u}} |H_u|^2 + {m^2_S}|S|^2 - (\lambda A_{\lambda} H_u H_d S + \text{ h.c.} ) +  (\frac{1}{3} \kappa A_{\kappa} S^3 + \text{ h.c.} ),
\label{eqn:PQlimit}
\end{align}
where $H_d$, $H_u$ and $S$ denote the neutral Higgs fields of the ${\bf H_d}$, ${\bf H_u}$ and ${\bf S}$ supermultiplets, respectively. For convenience, let's define its CP-even and CP-odd mass eigenstates as $H_i$, $i=1,2,3$, and $A_j$, $j=1,2$, respectively.

In contrast with the 2HDM case, the light CP-odd Higgs boson in the NMSSM often results from breaking an approximate global symmetry spontaneously, serving as an axion or a pseudo-Goldstone boson. Its appearance is thus less ``artificial''. Let us start with the tree-level mass matrix of the CP-odd Higgs bosons in the NMSSM: 
\begin{eqnarray}
\mathcal{M}^2_{P} & = & \left(
\begin{array}{cc}
m_A^2
&\lambda  v \left(\frac{m_A^2 }{2\mu }\sin 2 \beta -\frac{3 \kappa  \mu }{\lambda }\right)
\\
&
\lambda ^2 v^2 s_{2\beta}  \left(\frac{m_A^2}{4\mu ^2} \sin 2 \beta+\frac{3 \kappa }{2\lambda }\right)-\frac{3
   \kappa  A_{\kappa } \mu }{\lambda }\\
\end{array} 
\right), \ \  m_A^2 = \frac{2 \mu (A_\lambda + \kappa s) }{\sin 2 \beta} \nonumber\\
\end{eqnarray}
which yields a determinant
\begin{eqnarray}
\det(\mathcal{M}^2_{P}) = 9 \kappa \lambda v^2 \mu A_\lambda - \frac{6 A_\kappa \kappa \mu^2}{\lambda \sin2\beta} \left( A_\lambda + \frac{\kappa \mu }{\lambda} \right).
\end{eqnarray}
Necessarily, the scenarios with a light $A_1$ ($A_1$ denotes the lightest CP-odd Higgs boson) or $m_{A_1} \to 0$ yield $\det(\mathcal{M}^2_{P})  \to 0$
and viceversa, if such a stable vacuum exists. Among various possibilities, two have been studied extensively: 
R-symmetry (or R-limit) and Peccei-Quinn (PQ) symmetry (or PQ-limit), both of which yield a vanishing determinant at tree level. 
Another difference between these two class of scenarios is that the light CP-odd Higgs boson in the NMSSM is typically singlet-like. This can be understood since the Goldstone boson of a spontaneously-broken global $U(1)$ symmetry is manifested as  
\begin{eqnarray}
A_1 \sim \sum_i \frac{q_i v_i}{v_{U(1)}} \Phi_i, \ \ \Phi_i = S, H_u, H_d.
\end{eqnarray}
Here $v_{U(1)} = \sqrt{\sum q_i^2 v_i^2}$ is the $U(1)$ breaking scale and $q_i$ is the $U(1)$ charge of $\Phi_i$. An effective parameter $\mu = \lambda \langle v_S\rangle $ of the electroweak scale with  $\lambda \sim \mathcal O(0.1)$ naturally yields $v_S \gg v_u, v_d$, and hence a singlet-like pseudo-Goldstone boson. This feature renders such a light boson much more difficult to probe at colliders, compared to the 2HDM case. Next, we will evaluate the collider constraints on 
these two scenarios. 

\begin{enumerate}

\item R-limit: $A_\lambda \to 0$, $A_\kappa \to 0$, where the theory is approximately invariant under the transformation 
\begin{eqnarray}
H_u \to H_u \exp(i \phi_{\rm R}), \ \ H_d \to H_d \exp(i \phi_{\rm R}), \ \ S \to S \exp(i \phi_{\rm R}), 
\end{eqnarray}
and the tree-level couplings of the R-axion $A_1$ with the top and bottom quarks are given by  
\begin{eqnarray}
y_{A_1 tt} = \frac{2 \lambda v \cos^2 \beta}{\mu}, \ \  y_{A_1 bb} = \frac{2 \lambda v \sin^2 \beta}{\mu}.
\end{eqnarray}
In this scenario, both $\lambda$ and $\kappa$ can be large, yielding a sizeable contribution to the mass of the SM-like Higgs boson at tree level. Hence, a large value for $\tan\beta$ is unnecessary. A scan in the parameter space in this scenario is performed using NMSSMTools 4.2.1~\cite{NMSSMTools} including all built-in constraints, such as from Higgs searches, superpartner searches, muon $g-2$, flavour physics, invisible $Z$-decay, and the constraints from $\Upsilon$ decays (with the exception of the Landau pole test and DM related-constraints, which are not considered). 
The resulting values for the $y_{A_1tt}$ and $y_{A_1bb}$ couplings are compared to the expected collider bounds in Fig.~\ref{fig:reach_RS}(a).
Depending on the parameter values, the magnitude of $y_{A_1tt}$ in this scenario can be up to $\sim 0.5$. Only for an integrated luminosity of 3000~fb$^{-1}$ the LHC can probe a coupling $y_{A_1tt}$ as small as $~0.5$ via the $t\bar{t} A_1$, $A_1 \to b\bar{b}$ channel. Therefore this scenario is difficult to probe, even at the HL-LHC. 

\begin{figure}[htbp]
\begin{center}
\begin{tabular}{cc}
\includegraphics[width=0.45\textwidth,natwidth=610,natheight=642]{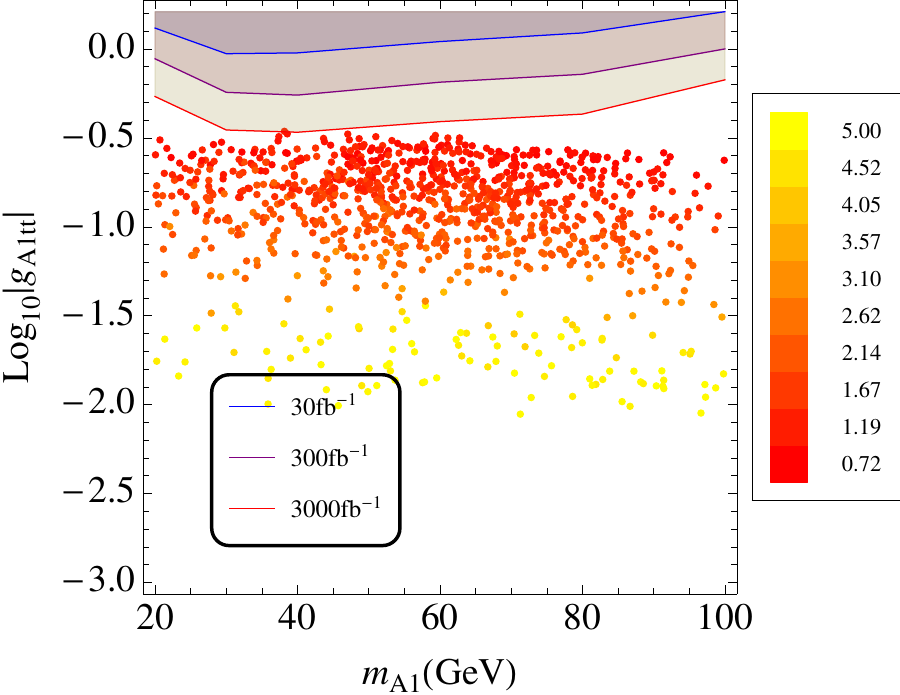} &
\includegraphics[width=0.45\textwidth,natwidth=610,natheight=642]{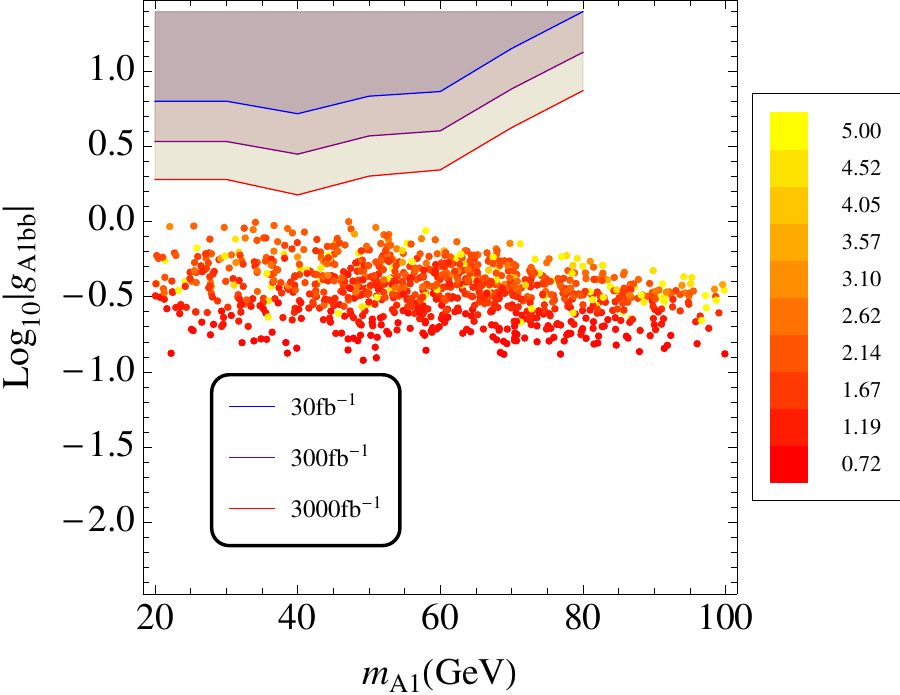} \\
(a) & (b) \\
\end{tabular}
\caption{Sensitivity reach (at 95\% CL)  to the R-limit scenario in the NMSSM via the (a) $t\bar{t}A_1$ and (b) $b\bar{b} A_1$ channels. The scan is over all
  parameters, in the ranges $0.1 \leq \lambda \leq 0.6$, $0.1 \leq
  \kappa \leq 0.6$, $-7 \leq A_\lambda \leq 7$~GeV, $-7 \leq A_\kappa \leq 0$~GeV,
  $0.1 \leq \tan \beta \leq 5$, and $100 \leq \mu \leq 500$~GeV.  We
  have assumed soft squark masses of 2 TeV, slepton masses of 200 GeV,
  $A_{u,d,e} = -3.5$ TeV, and bino, wino and gluino masses of
  100, 200, and 2000 GeV, respectively. The hue of the scatter points represents the correspondent $\tan\beta$ values.}
\label{fig:reach_RS} 
\end{center}
\end{figure}

\item PQ-limit: $\frac{\kappa}{\lambda} \to 0$, $A_\kappa \to 0$, where the theory is approximately invariant under the transformation 
\begin{eqnarray}
H_u \to H_u \exp(i \phi_{\rm PQ}), \ \ H_d \to H_d \exp(i \phi_{\rm PQ}), \ \ S \to S \exp(-2i \phi_{\rm PQ}), 
\end{eqnarray}
and the tree-level mass of the PQ pseudo-Goldstone boson $A_1$ are given given by  
\begin{eqnarray}
m_{A_1} = - \frac{3 \kappa A_\kappa \mu}{\lambda}.
\end{eqnarray}
This scenario has been proposed as a supersymmetric benchmark for sub-electroweak scale (singlino-like) DM~\cite{Draper:2010ew}, since its lightest neutralino is generically singlino-like and lighter than the electroweak scale. Particularly, in this scenario $A_1$ can serve as the mediator for  DM annihilation into a bottom quark pair and explain the diffuse gamma-ray excess from the Galactic Centre~\cite{Huang:2014cla,Cheung:2014lqa}. In this limit, the tree-level couplings of $A_1$ with the top and bottom quarks are given by 
\begin{eqnarray}
y_{A_1 tt} = \frac{\lambda v \cos^2 \beta}{\mu}, \ \ y_{A_1 bb} = \frac{\lambda v \sin^2 \beta}{\mu},
\end{eqnarray}
and so smaller by a factor of two than the corresponding couplings in the R-limit. Furthermore, a smaller $\lambda$ is favoured in this limit and a relatively large $\tan\beta$ is needed to generate a mass of 125 GeV for the SM-like Higgs boson. Therefore, the coupling $y_{A_1 tt}$ tends to be smaller than in the R-limit scenario. 
The resulting values for the $y_{A_1tt}$ and $y_{A_1bb}$ couplings are compared to the expected collider bounds in Fig.~\ref{fig:reach_RS}(b).
For most of the points, the magnitude of $y_{A_1 tt}$ is below 0.1, which renders this scenario extremely difficult to probe at the using the $t\bar tA_1$ channel.\footnote{For alternative way to probing this scenario, using exotic Higgs decays, see e.g. Ref.~\cite{Huang:2013ima,Huang:2014cla,Butter:2015fqa}.}
 
\begin{figure}[htbp]
\begin{center}
\begin{tabular}{cc}
\includegraphics[width=0.45\textwidth,natwidth=610,natheight=642]{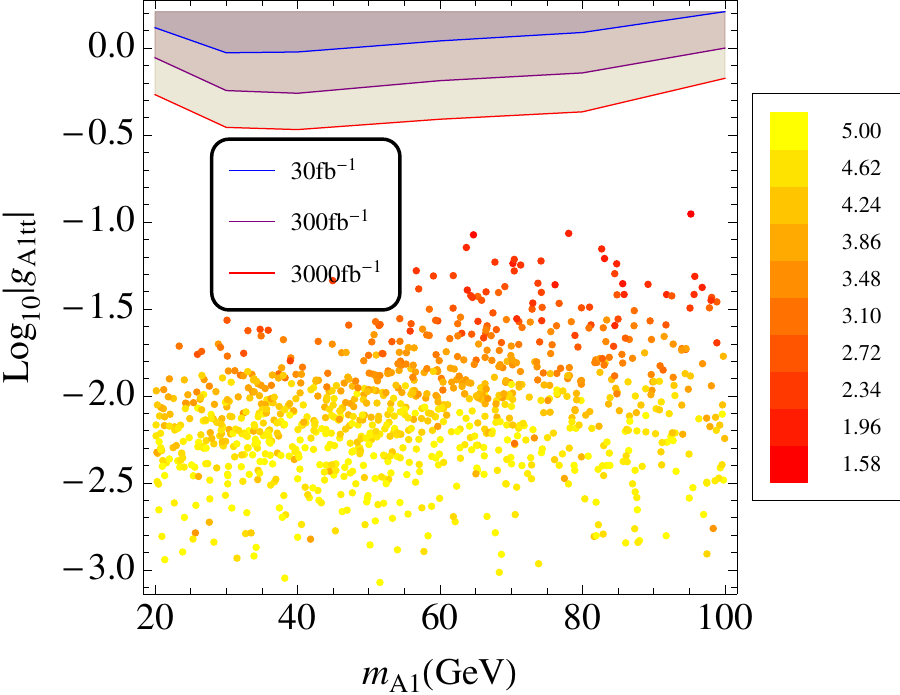} &
\includegraphics[width=0.45\textwidth,natwidth=610,natheight=642]{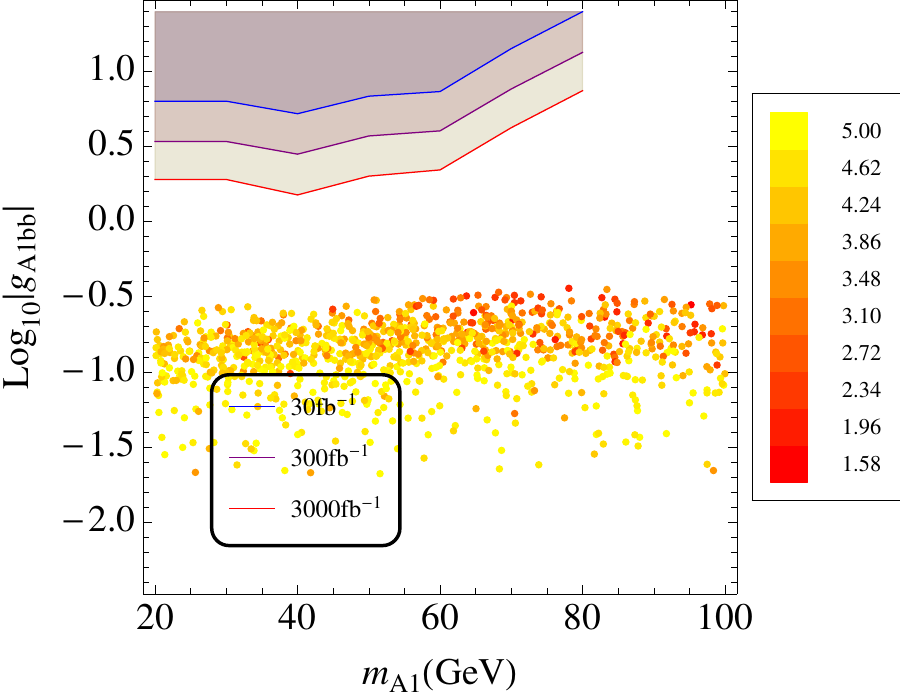} \\
(a) & (b) \\
\end{tabular}
\caption{Sensitivity reach (at 95\% CL)  to the PQ-limit scenario in the NMSSM via the (a) $t\bar{t}A_1$ and (b) $b\bar{b} A_1$ channels. The scan is over all
  parameters, in the ranges $0.06 \leq \lambda \leq 0.6$, $5 \leq
  \kappa / \lambda \leq 100$, $|\varepsilon'| = \left| A_\lambda/\mu \tan\beta -1\right| \leq 0.25$, $-100 \leq A_\kappa \leq 0$~GeV,
  $0.1 \leq \tan \beta \leq 5$, and $100 \leq \mu \leq 500$~GeV.  We
  have assumed soft squark masses of 2 TeV, slepton masses of 200 GeV,
  $A_{u,d,e} = 3.5$ TeV, and bino, wino and gluino masses of
  100, 200, and 2000 GeV, respectively. The hue of the scatter points represents the correspondent $\tan\beta$ values.}
\label{fig:reach_PQ} 
\end{center}
\end{figure}

\end{enumerate}

Finally, we stress that the $b\bar b A_1$ channel doesn't help much in probing the R- and PQ-limit scenarios. The sensitivities of both searches are suppressed by the mixture with the singlet. Even worse, the mixing is approximately $\tan\beta$ enhanced, further suppressing the sensitivity of the $b\bar b A_1$ in probing the large $\tan\beta$ region in both scenarios.


\section{Conclusions}
\label{sec:conclusion}

Searches for CP-odd scalars, as predicted by many extensions of the Standard Model and motivated by some recent astroparticle observations, are part of the core program of upcoming LHC runs at $\sqrt{s}=13$ and 14 TeV. Searches at LEP and during Run 1 of the LHC at $\sqrt{s}=7$ and 8 TeV have placed only weak constraints on the coupling strengths of CP-odd scalars with top and bottom quarks, or in their allowed mass range.

Using a simplified model approach for the signal, we have carried out a detailed study to evaluate the prospects at the LHC for probing scenarios with a 
CP-odd scalar with mass $20 \leq m_A < 100$ GeV, via the process $pp \to t\bar{t}A$ with subsequent decay $A\to b\bar{b}$. To separate the signal from the large background from $t\bar{t}$+jets production, we apply jet substructure techniques, reconstructing the mass of the CP-odd scalar as the mass of a 
large-radius jet containing two $b$-tagged subjets. 
The chosen method allows for a so-called 'bump hunt' over a fairly smooth background, and it may be the most promising strategy for searching for a CP-odd scalar with mass $\lesssim 50$ GeV, i.e. about twice the typical minimum $\pt$ cut for narrow jets used in standard LHC searches. A significant effort has been made in developing a semi-realistic experimental analysis, including a fairly complete description of systematic uncertainties and the usage of sophisticated statistical tools to constrain  {\em in-situ} the effect of systematic uncertainties, thus limiting their impact on the search sensitivity. We then derive expected upper limits on the production cross section times branching ratio using the CL$_{\rm{s}}$ method.

In specific models, e.g. 2HDM or NMSSM, the coupling of the $A$ boson with the top quark is related to other couplings in a well-defined way. Hence, the upper 
limits obtained on this coupling for a given mass $m_A$, can be used to bound other couplings of these models indirectly or as input for 
a global coupling fit. We find that in a type-I and type-II 2HDM the LHC can constrain a large fraction of the $(m_A, \tan \beta)$ parameter space, including 
the region preferred to explain the diffuse gamma-ray excess from the Galactic Centre as dark-matter annihilation via a CP-odd scalar mediator and decaying
into $b\bar{b}$. 
However, in the case of the NMSSM with a light CP-odd scalar, a Goldstone boson of either a spontaneously-broken R- or PQ-symmetry, 
the LHC appears to have very limited sensitivity in probing these models. 

Hence, depending on the concrete embedding of the scalar sector into a UV-complete theory, the LHC can provide complementary information, not accessible at either indirect detection experiments or electron-positron colliders, on the existence of CP-odd scalars, their mass and couplings to third-generation fermions.

\section{Acknowledgments}
This research was supported in part by the European Commission through the 'HiggsTools' Initial Training Network PITN-GA-2012-316704 (M.S.) and 
by the Spanish Ministerio de Econom\'ia y Competitividad  under projects FPA2012-38713 and Centro de Excelencia Severo Ochoa SEV-2012-0234 (M.C., T.F. and A.J). T.L. is supported by his start-up fund at the HKUST. T.L.  would also like to thank Y. Jiang for useful discussions and acknowledge the hospitality of the Jockey Club Institute for Advanced Study, HKUST, where part of this work was completed.

\appendix
\label{app:amplitudes}

\bibliography{ref}
\bibliographystyle{ArXiv}

\end{document}